%% file: Main.tex
\pdfoutput=1
\documentclass[12pt,a4paper]{article}

\usepackage{ifthen} 
\newboolean{pdflatex}
\setboolean{pdflatex}{true} 

\newboolean{articletitles}
\setboolean{articletitles}{true} 

\newboolean{uprightparticles}
\setboolean{uprightparticles}{false} 

\def\paperauthors{LHCb collaboration} 
\def\paperasciititle{Measurement of exclusive J/psi and psi(2S) production at sqrt(s)=13 TeV} 
\def\papertitle{Measurement of exclusive\\ \jpsi and \psitwos production at $\sqrt{s}=13\tev$} 
\def\paperkeywords{{High Energy Physics}, {LHCb}} 
\def\papercopyright{\the\year\ CERN for the benefit of the LHCb collaboration} 
\def\paperlicence{CC BY 4.0 licence}
\def\paperlicenceurl{https://creativecommons.org/licenses/by/4.0/}

\newcommand{\removed}[1]{\xspace}
\newcommand{\added}[2][NONE]{\ifthenelse{\equal{#1}{NONE}}{}{\removed{#1}}{#2}\xspace}

\newif\ifEnableSectionTOCLinks
\EnableSectionTOCLinksfalse 

\input{preamble}
\usepackage{local}
\usepackage{booktabs}
\usepackage{xcolor}
\definecolor{LightGray}{gray}{0.9}
\usepackage{placeins}

\begin{document}

\renewcommand{\thefootnote}{\fnsymbol{footnote}}
\setcounter{footnote}{1}

\input{title-LHCb-PAPER}


\renewcommand{\thefootnote}{\arabic{footnote}}
\setcounter{footnote}{0}

\cleardoublepage


\pagestyle{plain} 
\setcounter{page}{1}
\pagenumbering{arabic}


\input{body}

\input{acknowledgements}

\newpage
\input{appendix}

\FloatBarrier
\addcontentsline{toc}{section}{References}
\bibliographystyle{LHCb}
\bibliography{main,standard,LHCb-PAPER,LHCb-CONF,LHCb-DP,LHCb-TDR,exp,theory}
 
\FloatBarrier
\newpage
\input{Authorship_LHCb-PAPER-2024-012}

\end{document}

%% file: preamble.tex
\usepackage[top=1in, bottom=1.25in, left=1in, right=1in]{geometry}

\usepackage{microtype}
\usepackage{lineno}  
\usepackage{xspace} 
\usepackage{caption} 

\usepackage{graphicx}  
\usepackage{color}
\usepackage{colortbl}
\graphicspath{{./figs/}} 

\usepackage{amsmath} 
\usepackage{amssymb}
\usepackage{amsfonts}
\usepackage{upgreek} 

\newcommand*\patchAmsMathEnvironmentForLineno[1]{%
\expandafter\let\csname old#1\expandafter\endcsname\csname #1\endcsname
\expandafter\let\csname oldend#1\expandafter\endcsname\csname
end#1\endcsname
 \renewenvironment{#1}%
   {\linenomath\csname old#1\endcsname}%
   {\csname oldend#1\endcsname\endlinenomath}%
}
\newcommand*\patchBothAmsMathEnvironmentsForLineno[1]{%
  \patchAmsMathEnvironmentForLineno{#1}%
  \patchAmsMathEnvironmentForLineno{#1*}%
}
\AtBeginDocument{%
\patchBothAmsMathEnvironmentsForLineno{equation}%
\patchBothAmsMathEnvironmentsForLineno{align}%
\patchBothAmsMathEnvironmentsForLineno{flalign}%
\patchBothAmsMathEnvironmentsForLineno{alignat}%
\patchBothAmsMathEnvironmentsForLineno{gather}%
\patchBothAmsMathEnvironmentsForLineno{multline}%
\patchBothAmsMathEnvironmentsForLineno{eqnarray}%
}

\usepackage{hyperxmp}

\usepackage[pdftex,
            pdfauthor={\paperauthors},
            pdftitle={\paperasciititle},
            pdfkeywords={\paperkeywords},
            pdfcopyright={Copyright (C) \papercopyright},
            pdflicenseurl={\paperlicenceurl}]{hyperref}

\usepackage[colorinlistoftodos,textsize=scriptsize]{todonotes}

\usepackage[bottom,flushmargin,hang,multiple]{footmisc}

\usepackage[all]{hypcap} 

\input{lhcb-symbols-def} 

\hypersetup{
  colorlinks   = true, 
  urlcolor     = blue, 
  linkcolor    = blue, 
  citecolor    = red   
}

\ifEnableSectionTOCLinks
    \usepackage[explicit]{titlesec} 
    
    \let\oldcontentsline\contentsline
    \renewcommand

    \titleformat{\section}{\normalfont\Large\bf}{\hyperlink{tocsection.\thesection}{{\thesection} \parbox[t]{\dimexpr\textwidth-1pc}{#1}}}{1pc}{}

    \titleformat{\subsection}{\normalfont\bf}{\hyperlink{tocsubsection.\thesubsection}{{\thesubsection} \parbox[t]{\dimexpr\textwidth-1pc}{#1}}}{1pc}{}

    \titleformat{name=\section,numberless}[display]{}{}{0pt}{\normalfont\Huge\bfseries #1}
\fi

\usepackage{cite} 
\usepackage{mciteplus}

%% file: lhcb-symbols-def.tex
\usepackage{xspace} 
\usepackage{upgreek}

\def\lhcb   {\mbox{LHCb}\xspace}

\def\velo   {VELO\xspace}

\def\herschel {\mbox{\textsc{HeRSCheL}}\xspace}
\def\MagUp {\mbox{\em Mag\kern -0.05em Up}\xspace}

\ifthenelse{\boolean{uprightparticles}}%
{
 
 \def\Pgamma      {\ensuremath{\upgamma}\xspace}

 \def\Pmu         {\ensuremath{\upmu}\xspace}

 \def\Ppi         {\ensuremath{\uppi}\xspace}

 \def\Pphi        {\ensuremath{\upphi}\xspace}                 
                  
 \def\Pchi        {\ensuremath{\upchi}\xspace}                 
 \def\Ppsi        {\ensuremath{\uppsi}\xspace}

 \def\PDelta      {\ensuremath{\Delta}\xspace}                 
 \def\PXi         {\ensuremath{\Xi}\xspace}                 
 \def\PLambda     {\ensuremath{\Lambda}\xspace}                 
 \def\PSigma      {\ensuremath{\Sigma}\xspace}                 
 \def\POmega      {\ensuremath{\Omega}\xspace}                 
 \def\PUpsilon    {\ensuremath{\Upsilon}\xspace}
 \let\oldPi\Pi
 \def\PPi         {\ensuremath{\oldPi}\xspace}

 \def\PB      {\ensuremath{\mathrm{B}}\xspace}                 
 \def\PD      {\ensuremath{\mathrm{D}}\xspace}                 
 \def\PJ      {\ensuremath{\mathrm{J}}\xspace}                 
 \def\PK      {\ensuremath{\mathrm{K}}\xspace}                 
 \def\Pb      {\ensuremath{\mathrm{b}}\xspace}                 
 \def\Pc      {\ensuremath{\mathrm{c}}\xspace}                 
                  
 \def\Pe      {\ensuremath{\mathrm{e}}\xspace}                 
 \def\Ps      {\ensuremath{\mathrm{s}}\xspace}

 \def\thebaroffset{0.0em}
}
{
 
 \def\Pgamma      {\ensuremath{\gamma}\xspace}

 \def\Pmu         {\ensuremath{\mu}\xspace}

 \def\Ppi         {\ensuremath{\pi}\xspace}

 \def\Pphi        {\ensuremath{\phi}\xspace}                 
                  
 \def\Pchi        {\ensuremath{\chi}\xspace}                 
 \def\Ppsi        {\ensuremath{\psi}\xspace}                 
                  
 \mathchardef\PDelta="7101
 \mathchardef\PXi="7104
 \mathchardef\PLambda="7103
 \mathchardef\PSigma="7106
 \mathchardef\POmega="710A
 \mathchardef\PUpsilon="7107
 \mathchardef\PPi="7105
 \def\PB      {\ensuremath{B}\xspace}                 
 \def\PD      {\ensuremath{D}\xspace}                 
 \def\PJ      {\ensuremath{J}\xspace}                 
 \def\PK      {\ensuremath{K}\xspace}                 
 \def\Pb      {\ensuremath{b}\xspace}                 
 \def\Pc      {\ensuremath{c}\xspace}                 
                  
 \def\Pe      {\ensuremath{e}\xspace}                 
 \def\Ps      {\ensuremath{s}\xspace}

 \def\thebaroffset{0.18em}
}
\newcommand{\offsetoverline}[2][\thebaroffset]{\kern #1\overline{\kern -#1 #2}}%

\makeatletter
\ifcase \@ptsize \relax
  \newcommand{\miniscule}{\@setfontsize\miniscule{4}{5}}
\or
  \newcommand{\miniscule}{\@setfontsize\miniscule{5}{6}}
\or
  \newcommand{\miniscule}{\@setfontsize\miniscule{5}{6}}
\fi
\makeatother

\DeclareRobustCommand{\optbar}[1]{\shortstack{{\miniscule (\rule[.5ex]{1.25em}{.18mm})}
  \\ [-.7ex] $#1$}}

\def\epem       {{\ensuremath{\Pe^+\Pe^-}}\xspace}

\def\mupm       {{\ensuremath{\Pmu^\pm}}\xspace} 
 
\def\mumu       {{\ensuremath{\Pmu^+\Pmu^-}}\xspace}

\def\ellell     {\ensuremath{\ell^+ \ell^-}\xspace}

\def\squark    {{\ensuremath{\Ps}}\xspace}

\def\cquark    {{\ensuremath{\Pc}}\xspace}
\def\cquarkbar {{\ensuremath{\overline \cquark}}\xspace}

\def\bquark    {{\ensuremath{\Pb}}\xspace}

\def\pion   {{\ensuremath{\Ppi}}\xspace}
\def\piz    {{\ensuremath{\pion^0}}\xspace}

\def\KorKbar {\kern \thebaroffset\optbar{\kern -\thebaroffset \PK}{}\xspace}

\def\D       {{\ensuremath{\PD}}\xspace}

\def\DorDbar {\kern \thebaroffset\optbar{\kern -\thebaroffset \PD}\xspace}

\def\Dp      {{\ensuremath{\D^+}}\xspace}
\def\Dm      {{\ensuremath{\D^-}}\xspace}

\def\DpDm    {\ensuremath{\Dp {\kern -0.16em \Dm}}\xspace}

\def\B       {{\ensuremath{\PB}}\xspace}

\def\BorBbar {\kern \thebaroffset\optbar{\kern -\thebaroffset \PB}\xspace}

\def\Bd      {{\ensuremath{\B^0}}\xspace}

\def\BdorBdbar {\kern \thebaroffset\optbar{\kern -\thebaroffset \Bd}\xspace}

\def\Bs      {{\ensuremath{\B^0_\squark}}\xspace}

\def\BsorBsbar {\kern \thebaroffset\optbar{\kern -\thebaroffset \Bs}\xspace}

\def\jpsi     {{\ensuremath{{\PJ\mskip -3mu/\mskip -2mu\Ppsi}}}\xspace}
\def\psitwos  {{\ensuremath{\Ppsi{(2S)}}}\xspace}

\def\chic     {{\ensuremath{\Pchi_\cquark}}\xspace}
\def\chiczero {{\ensuremath{\Pchi_{\cquark 0}}}\xspace}
\def\chicone  {{\ensuremath{\Pchi_{\cquark 1}}}\xspace}
\def\chictwo  {{\ensuremath{\Pchi_{\cquark 2}}}\xspace}
\def\chicJ    {{\ensuremath{\Pchi_{\cquark J}}}\xspace}
\def\Upsilonres  {{\ensuremath{\PUpsilon}}\xspace}
\def\Y#1S{\ensuremath{\PUpsilon{(#1S)}}\xspace}

\def\LorLbar     {\kern \thebaroffset\optbar{\kern -\thebaroffset \PLambda}\xspace}

\newcommand{\decay}[2]{\mbox{\ensuremath{#1\!\to #2}}\xspace} 

\def\to                 {\ensuremath{\rightarrow}\xspace}

\def\AT#1     {\ensuremath{A_{\mathrm{T}}^{#1}}\xspace}           

\def\C#1      {\ensuremath{\mathcal{C}_{#1}}\xspace}                       
\def\Cp#1     {\ensuremath{\mathcal{C}_{#1}^{'}}\xspace}                    
\def\Ceff#1   {\ensuremath{\mathcal{C}_{#1}^{\mathrm{(eff)}}}\xspace}        
\def\Cpeff#1  {\ensuremath{\mathcal{C}_{#1}^{'\mathrm{(eff)}}}\xspace}       
\def\Ope#1    {\ensuremath{\mathcal{O}_{#1}}\xspace}                       
\def\Opep#1   {\ensuremath{\mathcal{O}_{#1}^{'}}\xspace}                    

\newcommand{\aunit}[1]{\ensuremath{\text{\,#1}}}       

\newcommand{\tev}{\aunit{Te\kern -0.1em V}\xspace}
\newcommand{\gev}{\aunit{Ge\kern -0.1em V}\xspace}
\newcommand{\mev}{\aunit{Me\kern -0.1em V}\xspace}
\newcommand{\kev}{\aunit{ke\kern -0.1em V}\xspace}
\newcommand{\ev}{\aunit{e\kern -0.1em V}\xspace}
 
\newcommand{\mevc}{\ensuremath{\aunit{Me\kern -0.1em V\!/}c}\xspace}
\newcommand{\gevc}{\ensuremath{\aunit{Ge\kern -0.1em V\!/}c}\xspace}
\newcommand{\mevcc}{\ensuremath{\aunit{Me\kern -0.1em V\!/}c^2}\xspace}
\newcommand{\gevcc}{\ensuremath{\aunit{Ge\kern -0.1em V\!/}c^2}\xspace}
\newcommand{\gevgevcc}{\ensuremath{\gev^2\!/c^2}\xspace} 

\def\pb {\aunit{pb}\xspace}
\def\invpb {\ensuremath{\pb^{-1}}\xspace}
\def\fb   {\ensuremath{\aunit{fb}}\xspace}
\def\invfb   {\ensuremath{\fb^{-1}}\xspace}

\newcommand{\chisq}{\ensuremath{\chi^2}\xspace}

\def\gsim{{~\raise.15em\hbox{$>$}\kern-.85em
          \lower.35em\hbox{$\sim$}~}\xspace}
\def\lsim{{~\raise.15em\hbox{$<$}\kern-.85em
          \lower.35em\hbox{$\sim$}~}\xspace}

\def\sqs   {\ensuremath{\protect\sqrt{s}}\xspace}
\def\sqsnn {\ensuremath{\protect\sqrt{s_{\scriptscriptstyle\text{NN}}}}\xspace}
\def\pt         {\ensuremath{p_{\mathrm{T}}}\xspace}
\def\ptsq       {\ensuremath{p_{\mathrm{T}}^2}\xspace}
\def\ptot       {\ensuremath{p}\xspace}

\newcommand{\lum} {\ensuremath{\mathcal{L}}\xspace}

\def\evtgen     {\mbox{\textsc{EvtGen}}\xspace}

\def\geant      {\mbox{\textsc{Geant4}}\xspace}

\def\photos     {\mbox{\textsc{Photos}}\xspace}

\def\tell1  {TELL1\xspace}
\def\ukl1   {UKL1\xspace}

\newcommand{\eg}{\mbox{\itshape e.g.}\xspace}
\newcommand{\ie}{\mbox{\itshape i.e.}\xspace}
\newcommand{\etal}{\mbox{\itshape et al.}\xspace}

\newcommand{\phz}{\phantom{0}}
\newcommand{\lhcborcid}[1]{\href{https://orcid.org/#1}{\hspace*{0.1em}\raisebox{-0.45ex}{\includegraphics[width=1em]{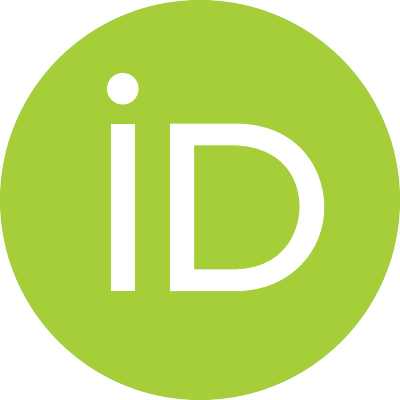}}}}


%% file: title-LHCb-PAPER.tex

\begin{titlepage}
\pagenumbering{roman}

\vspace*{-1.5cm}
\centerline{\large EUROPEAN ORGANIZATION FOR NUCLEAR RESEARCH (CERN)}
\vspace*{1.5cm}
\noindent
\begin{tabular*}{\linewidth}{lc@{\extracolsep{\fill}}r@{\extracolsep{0pt}}}
\ifthenelse{\boolean{pdflatex}}
{\vspace*{-1.5cm}\mbox{\!\!\!\includegraphics[width=.14\textwidth]{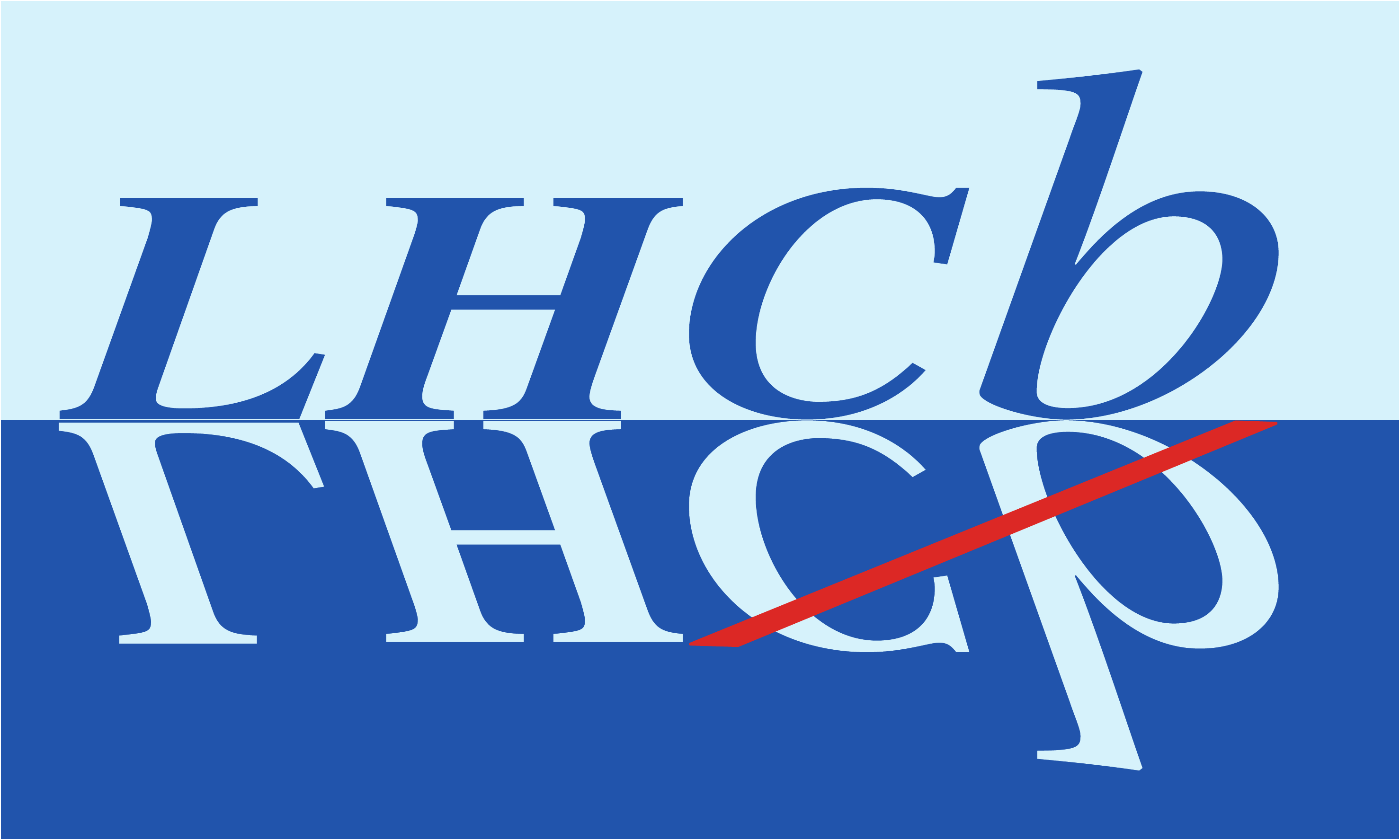}} & &}%
{\vspace*{-1.2cm}\mbox{\!\!\!\includegraphics[width=.12\textwidth]{figs/lhcb-logo.eps}} & &}%
\\
 & & CERN-EP-2024-213 \\  
 & & LHCb-PAPER-2024-012 \\  
 & & 28 February 2025 \\ 
\end{tabular*}

\vspace*{2.0cm}

{\normalfont\bfseries\boldmath\huge
\begin{center}
  \papertitle 
\end{center}
}

\vspace*{1.5cm}

\begin{center}
\paperauthors\footnote{Authors are listed at the end of this paper.}
\end{center}

\vspace{\fill}

\begin{abstract}
  \noindent  
  Measurements are presented of the cross-section for the central exclusive production of $J/\psi\to\mu^+\mu^-$ and $\psi(2S)\to\mu^+\mu^-$ processes in proton-proton collisions at $\sqrt{s} = 13 \tev$ with 2016--2018 data.
  They are performed by requiring both muons to be in the LHCb acceptance (with pseudorapidity  $2<\eta_{\mupm} < 4.5$) and  mesons in the rapidity range $2.0 < y < 4.5$. The integrated cross-section results are
  \begin{align*}
    \sigma_{J/\psi\to\mu^+\mu^-}(2.0<y_{J/\psi}<4.5,2.0<\eta_{\mu^{\pm}} < 4.5) &= 400 \pm  2 \pm 5 \pm 12 \pb\,,\\
    \sigma_{\psi(2S)\to\mu^+\mu^-}(2.0<y_{\psi(2S)}<4.5,2.0<\eta_{\mu^{\pm}} < 4.5) &= 9.40 \pm  0.15 \pm 0.13 \pm 0.27 \pb\,,
\end{align*}
        where the uncertainties are statistical, systematic and due to the luminosity determination.
        In addition, a measurement of the ratio of $\psi(2S)$ and $J/\psi$ cross-sections, at an average photon-proton centre-of-mass energy of $1\tev$, is performed, giving
\begin{equation*}
	\frac{\sigma_{\psi(2S)}}{\sigma_{J/\psi}} = 0.1763 \pm 0.0029 \pm 0.0008 \pm 0.0039 \,,
\end{equation*}
where the first uncertainty is statistical, the second systematic and the third due to the knowledge of the involved branching fractions.

For the first time, the dependence of the $J/\psi$ and $\psi(2S)$ cross-sections on the total transverse momentum transfer
is determined in $pp$ collisions
and is found consistent with the behaviour observed in electron-proton collisions.

\end{abstract}

\vspace*{2.0cm}

\begin{center}
  Published in
  sciPost Physics 18 (2025) 071
\end{center}

\vspace{\fill}

{\footnotesize 
\centerline{\copyright~\papercopyright. \href{\paperlicenceurl}{\paperlicence}.}}
\vspace*{2mm}

\end{titlepage}


\newpage
\setcounter{page}{2}
\mbox{~}
%

%% file: body.tex
\section{Introduction}
\label{sec:Introduction}

Deep inelastic scattering of leptons off protons provided the first proof that hadrons are not elementary but rather composed
of quarks~\cite{Bloom:1969kc,Breidenbach:1969kd}.
It is an essential tool to determine parton distribution functions (PDFs) inside protons, which are required to
make cross-section predictions at hadron colliders.
However, charged leptons interact electromagnetically and only probe the density of the quarks, which are charged.
The densities of the neutral gluons must be inferred, which can be done by studying how the quark PDFs evolve with the scale 
set by the mass of the exchanged virtual photon.
These PDFs are determined in fits~\cite{Hou:2019efy,NNPDF:2017mvq,Bailey:2020ooq} to multiple measurements,
including notably $e^\pm p$ scattering\removed{ at 
HERA}~\cite{H1:2015ubc,H1:2018flt}, and \added[LHCb measurements]{forward production} of 
vector bosons~\cite{LHCb-PAPER-2015-001,LHCb-PAPER-2015-003,LHCb-PAPER-2015-049,LHCb-PAPER-2012-008}
and heavy-quarks~\cite{LHCb-PAPER-2015-041,Gauld:2016kpd,PROSA:2015yid,Bertone:2018dse} \added{in $pp$ collisions}.
Due to a lack of data at low $x$, the fraction of hadron momentum carried by the parton,
the uncertainties attributed to the gluon PDFs are large at low $x$ and are even
compatible with an unphysical decrease of the gluon density with $x$~\cite{Flett:2022stc}.
Other methods are thus required to access the gluonic PDF.

\begin{figure}[b]\centering\def\hh{0.14\textheight}
\includegraphics[height=\hh]{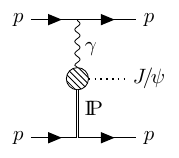}
\includegraphics[height=\hh]{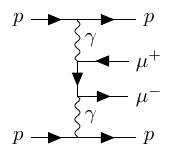}
\includegraphics[height=\hh]{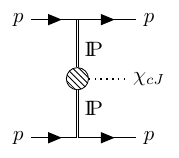}
\includegraphics[height=\hh]{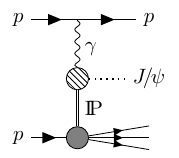}
	\caption{Feynman diagrams for 
 signal and background processes. From left to right: signal CEP $J\mskip -3mu/\mskip -2mu\psi$ photoproduction, where ${\rm I\!P}$ stands for a colourless superposition of gluons, sometimes referred to as a pomeron; continuum dimuon production; exclusive $\chi_c$ ($J=0,1,2$) production via double pomeron exchange;  inelastic $pp$ collision where a proton dissociates.}\label{fig:cep-process}\label{fig:feynmannD}  
\end{figure}

Central exclusive vector-meson production (CEP) in $pp$ collisions
is the quasi-elastic production of a single meson, leaving the protons intact.
Exclusive charmonium production results from the conversion of a virtual photon close to its mass shell into a \cquark\cquarkbar pair, which hadronises into a \jpsi  or \psitwos meson. 
These processes probe the gluonic PDF at the scale of the charm quark mass.
The exclusivity of the process requires that, at leading order, two gluons are exchanged with the target hadron.
Thus the cross-section approximately scales as gluon density squared~\cite{Ryskin:1992ui,Mantysaari:2022kdm,Flett:2019pux,Diehl:2003ny}.
The process and the main backgrounds are depicted in Fig.~\ref{fig:cep-process}.

Exclusive scattering processes also give access to the total transverse momentum
transfer $\Delta t$, the square of the difference between the momenta of the incoming and outgoing proton, which is a Fourier conjugate to the impact parameter between two colliding hadrons.
As such, the $\Delta t$ spectra are sensitive to the spatial
distribution of colour charge~\cite{Klein:2019qfb}.
Several predictions (see \eg Ref.~\cite{Mantysaari:2022kdm}) calculate the exclusive production cross-section
at $\Delta t\sim0$.
The cross-section falls exponentially versus $\Delta t$ with a slope determined experimentally, 
which can be used to infer the total exclusive cross-section.
In the present paper, this slope is determined in ten intervals of rapidity for the \jpsi and \psitwos
mesons.
As the outgoing protons are not detected at LHCb, $\Delta t$ is not directly accessible and the transverse momentum squared, \ptsq, of the charmonium state is used as a proxy.

The photoproduction cross-section of a charmonium state is sensitive to the radial wave function of the charmonium state in a region where the \psitwos wave function has a radial node but the \jpsi wave function does not.
As a result, the \psitwos photoproduction cross-section is expected to be suppressed with respect to that of \jpsi mesons~\cite{Nemchik:1997xb,Nemchik:2000de,Hentschinski:2020yfm,Hufner:2000jb,Kopeliovich:1991pu, Kopeliovich:1993pw,Nemchik:1994fp,Armesto:2014sma,Lappi:2010dd,Kowalski:2006hc}.
With many theoretical uncertainties cancelling, predictions for the ratio of $\psitwos$ and $\jpsi$ cross-sections can be determined more precisely than the individual cross-sections.

Exclusive \jpsi and \psitwos production in $pp$ collisions at the LHC have previously been measured at centre-of-mass energies of $\sqrt{s}=7$\tev~\cite{LHCb-PAPER-2012-044,LHCb-PAPER-2013-059}
and 13\tev~\cite{LHCb-PAPER-2018-011}.
Exclusive double-charmonium~\cite{LHCb-PAPER-2014-027} and \Upsilonres~\cite{LHCb-PAPER-2015-011} production have been measured at 7 and 8\tev,
and that of \jpsi{}\Pphi at 13\tev~\cite{LHCb-PAPER-2023-043}.
Charmonia production has also been studied in ultra-peripheral $p$Pb~\cite{ALICE:2023mfc} and
PbPb~\cite{ALICE:2019tqa,ALICE:2023svb,LHCb-PAPER-2022-012,CMS:2023snh} collisions. 

The previous LHCb measurements have been used to update PDF fits~\cite{Flett:2019pux,Boer:2023mip},
and thus improve predictions of \jpsi and \Upsilonres CEP cross-sections~\cite{Mantysaari:2022kdm,Flett:2021fvo,Flett:2020duk};
make predictions~\cite{Sharma:2023njj,Mantysaari:2023xcu,Mantysaari:2022sux,Flett:2022ues,Guzey:2020ntc,Henkels:2020kju,Cao:2018hvy} for ultra-peripheral photoproduction processes at RHIC~\cite{STAR:2023nos,STAR:2023gpk} and the LHC~\cite{LHCb-PAPER-2022-012,ALICE:2021gpt,ALICE:2023jgu,CMS:2023snh};
determine the meson-proton scattering length~\cite{Wang:2022xpw} and 
extract the proton mass radius from the \jpsi and \psitwos cross-sections~\cite{Wang:2022vhr}.
Based on these cross-sections, Ref.~\cite{ArroyoGarcia:2019cfl}
claims that LHCb data show evidence of gluon saturation, 
\ie the slowing down of the growth of gluon densities as $x$ decreases
due to gluon emission and recombination balancing each other, 
while the authors of Ref.~\cite{Flett:2020duk} disagree.
Such effects would usually be expected in heavy-ion collisions. 

This paper presents a measurement of exclusive \jpsi and \psitwos production in proton-proton collisions at \mbox{$\sqs=13\tev$} in the forward direction,
in ten intervals of rapidity between 2.0 and 4.5.
The data used were collected with the LHCb detector at the LHC between 2016 and 2018,
corresponding to an integrated luminosity of 4.4\invfb, which is
twenty times larger than that used in Ref.~\cite{LHCb-PAPER-2018-011}.
This larger sample permits \added{a better control of background shapes, implemented in a two-dimensional fit in dimuon mass and transverse-momentum squared.
For the first time, a} measurement of the \psitwos cross-section in the same
rapidity intervals as for the \jpsi cross-section, and thus the determination of their ratio as a function of rapidity\added{~is presented}.
\section{Detector, simulation and data sample}
\label{sec:Detector}

The \lhcb detector~\cite{LHCb-DP-2008-001,LHCb-DP-2014-002} is a single-arm forward
spectrometer covering the \mbox{pseudorapidity} range $2<\eta <5$,
designed for the study of particles containing \bquark or \cquark
quarks. The detector includes a high-precision tracking system
consisting of a silicon-strip vertex detector (VELO) surrounding the $pp$
interaction region~\cite{LHCb-DP-2014-001}, a large-area silicon-strip detector located
upstream of a dipole magnet with a bending power of about
$4{\mathrm{\,T\,m}}$, and three stations of silicon-strip detectors and straw
drift tubes~\cite{LHCb-DP-2017-001}
placed downstream of the magnet.
The tracking system provides a measurement of the momentum, \ptot, of charged particles with
a relative uncertainty that varies from 0.5\% at low momentum to 1.0\% at 200\gevc.
Photons
are identified by a calorimeter system consisting of
scintillating-pad and preshower detectors (SPD), and electromagnetic
and hadronic calorimeters. Muons are identified by a
system composed of alternating layers of iron and multiwire
proportional chambers~\cite{LHCb-DP-2012-002}.

The pseudorapidity coverage of the LHCb detector is extended by the \herschel system, composed of forward shower counters consisting of five
planes of scintillators with three planes at 114, 19.7 and 7.5\:m upstream of the interaction
point, and two downstream at 20 and 114\:m. At each location, there are four quadrants
of scintillators, whose information is recorded in every beam crossing by photomultiplier
tubes, giving a total of 20 channels in \herschel~\cite{LHCb-DP-2016-003}.
These are calibrated using data taken
without beams circulating at the end of each LHC fill~\cite{SanchezGras:2842971}.
The pseudorapidity ranges covered
by the VELO and \herschel are different. For the VELO the region is $-3.5 < \eta < -1.5$ and
$2 < \eta < 5$, and for \herschel the region is $-10 < \eta < -5$ and $5 < \eta < 10$.

The online event selection is performed by a trigger~\cite{LHCb-DP-2012-004,LHCb-DP-2019-001}
that consists of a hardware stage, based on information from the calorimeter and muon
systems, followed by a software stage, which applies a full event
reconstruction. The distinct signature of CEP events is their low multiplicity.
Consequently, at the hardware stage, the trigger selects events
containing at least one muon with $\pt>192\mevc$ and fewer than 20 hits in the SPD
detector. At the software stage events are selected if they contain
two muons with $\pt>400\mevc$, fewer than 10 tracks in the VELO, of which
at most four are reconstructed in the backward direction~\cite{LHCb-DP-2014-001}.
A sample used for the determination of trigger, reconstruction
and particle identification (PID) efficiencies is collected
requiring a single muon with $\pt>500\mevc$ and the same multiplicity
requirements as for the default selection.

The data used were collected between July 2016 and October 2018.
The early 2016 data are not used as relevant trigger selections were
not yet included.
Data from the last month of data taking in 2018 is also discarded as it was affected
by a noisy SPD readout board, which biases the number of SPD hits in
low-multiplicity events.

Offline, events are required to contain only the two muon candidates,
which should be of good quality\added{~\cite{LHCb-DP-2013-002}}, and identified as such\added{~\cite{LHCb-DP-2013-001}}, which
implies that their momentum exceeds $3\gevc$, the threshold to cross the
calorimeter and reach the muon system.
The event should contain no additional tracks in the VELO, and no photons
other than those that are consistent with being radiated from the passage of
muons through the detector material.

The muons from CEP signal \jpsi decays are well outside of the \herschel acceptance;
these counters are used to veto charged particles from the proton dissociating.
The CEP cross-section measurements are performed with events that contain no
such additional particles,
\ie \herschel signals consistent with noise.
The remaining events are retained for background studies.
The \herschel response is described using a discriminating \chisq-like
variable that quantifies the activity above noise taking into
account correlations between the counters\added{~\cite{LHCb-DP-2016-003}}.
The selection requirement is optimised using low-mass low-\ptsq dimuon
pairs, which are dominated by two-photon fusion.

Simulation is required to model the effects of the detector acceptance and the
imposed selection requirements, and to study specific backgrounds.
In the simulation, the charmonium candidate is generated and decayed using
{\tt SuperChic2}~\cite{Harland-Lang:2015cta}, with the exception
of \decay{\psitwos}{\jpsi X} processes (where $X$ is any combination of particles, mostly \pion\pion),
for which the decay is handled
by \evtgen~\cite{Lange:2001uf}.
Final-state radiation is generated using \photos~\cite{davidson2015photos}.
The interaction of the generated particles with the detector, and its response,
are implemented using the \geant
toolkit~\cite{Allison:2006ve, *Agostinelli:2002hh} as described in
Ref.~\cite{LHCb-PROC-2011-006}.
The \textsc{ROOT}~\cite{Brun:1997pa} and
\lhcb~\cite{Corti:2006yx,Koppenburg:2006kk,Tsaregorodtsev:2010zz} software frameworks are
used for the initial data preparation,
while the analysis is written in the \textsc{Python} language with standard scientific
packages~\cite{Harris:2020xlr,gough2009gnu,Verkerke:2003ir,James:1975dr,Hunter:2007ouj,Snakemake}.

The total integrated luminosity of the used data sample is
determined
using empty-event counters calibrated by van der Meer scans and beam profile
measurements~\cite{LHCb-PAPER-2014-047} and is found to be $\intlumi=4.41\pm0.13\invfb$.
Due to the multiplicity requirements imposed in the trigger and
the offline selection, only events with a single $pp$ interaction are selected.
The useful integrated luminosity is thus reduced by the fraction
of events with a single interaction containing at least two \velo tracks.
The number of such visible $pp$ interactions per beam crossing, $n$, is assumed to follow a
Poisson distribution, $P(n) = \mu^n e^{-\mu}/n!$, with mean $\mu$.
The fraction of useful integrated luminosity, \effintlumi, corresponding
to events with $n=1$, is given by
\begin{equation}\label{eq:fracLumi}
	f_{\lum} = \frac{\effintlumi}{\intlumi} = \frac{P(n=1)}{\sum\limits_{n = 0}^{\infty} n P(n)} = \frac{\mu e^{-\mu}}{\sum\limits_{n = 0}^{\infty} n \frac{\mu^n e^{-\mu}}{n!}} = e^{-\mu}.
\end{equation}
The value of $\mu$ depends on running conditions and it is determined in periods of up to one hour of stable running conditions~\cite{LHCb-PAPER-2014-047}.
In most running periods $\mu$ is close to 1.1,
with variations of less than 10\%, corresponding to an average $f_{\lum}\simeq0.33$.
The corresponding useful
integrated luminosity is $\effintlumi=1522\pm44\invpb$, where the uncertainty is dominated
by that on \intlumi.

\section{Two-dimensional signal fits}\label{sec:Fit}

The primary challenge in this analysis is separation of the elastic CEP and
inelastic proton-dissociation (PD) components, shown in Fig.~\ref{fig:feynmannD}.
The latter consists of events where the proton dissociates, producing
charged particles in the very forward acceptance.
These are vetoed by the \herschel requirement, which however is
not perfect and thus leaves some PD backgrounds in the selected signal sample.
The different \ptsq distributions of PD and CEP charmonia are therefore also
exploited. 
The properties of the PD component are determined from a control sample \added{that is free from any CEP signal contribution. This sample is} obtained by inverting the
\herschel veto, and requiring $0.9<\ptsq<5.0\gevgevcc$, where the CEP contribution,
which populates the low-\ptsq region, is negligible.

Two other backgrounds are accounted for: QED continuum dimuon production
and \jpsi feed-down from higher-mass charmonia,
namely \psitwos, and $\chicJ(1P)$ ($J=0,1,2)$, referred to as \chic below unless otherwise specified.
Other feed-down contributions, such as those from \Upsilonres resonances or \bquark hadrons,
are negligible because of the \velo-tracks veto.

\begin{figure}[!t]
	\centering
        \includegraphics[width=0.48\textwidth]{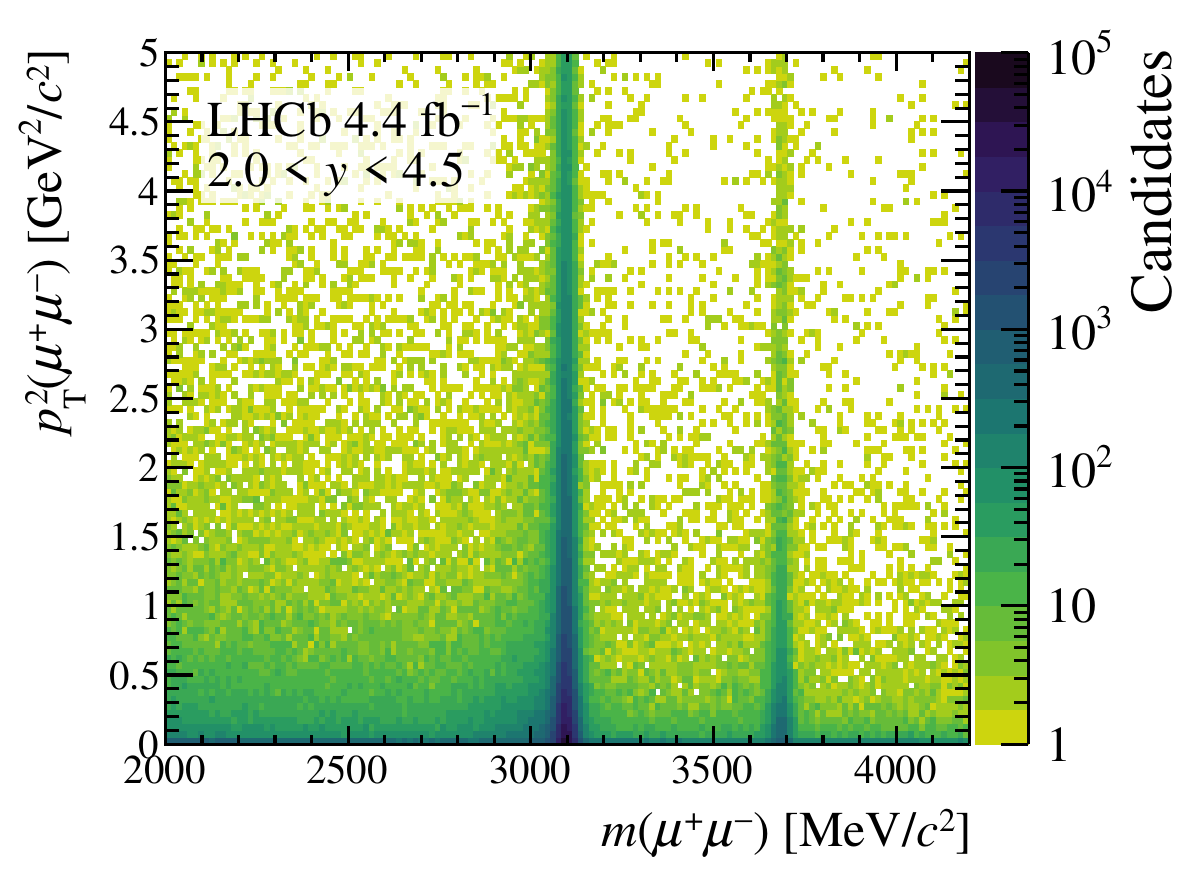}\hskip 0.03\textwidth%
        \includegraphics[width=0.48\textwidth]{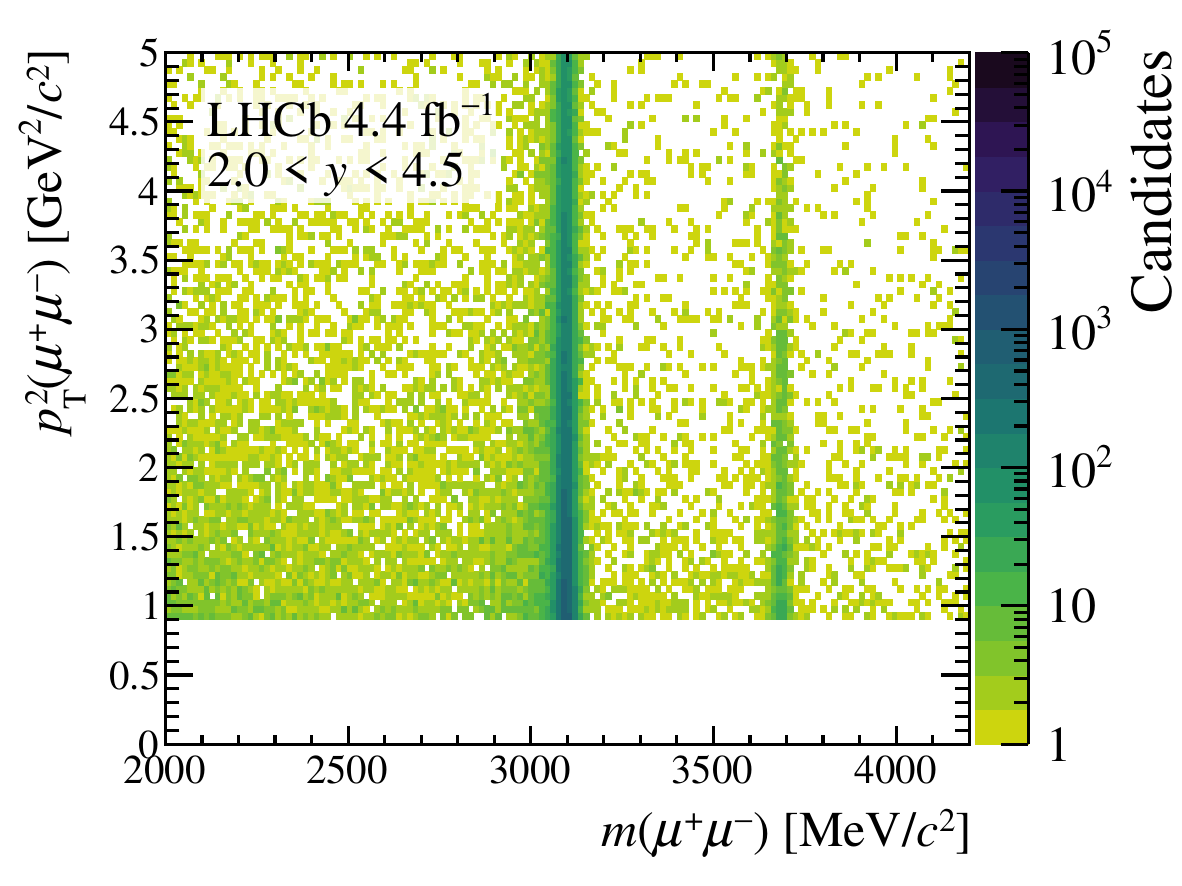}
	\caption{Two-dimensional mass-$p_{\mathrm{T}}^2$ distributions for the (left) signal and (right)
          control samples. }
	\label{fig:2DmassPTsq-colz}
\end{figure}

As there is a correlation between the dimuon mass and \ptsq distributions for non-peaking backgrounds, \added{the data are fit by} a two-dimensional
\added[fit to the data]{model in mass and \ptsq} in each of ten rapidity intervals\removed{ is performed}. The considered regions are
$2000<m_{\mumu}<4200\mevcc$ and $\ptsq<5\gevgevcc$.
The relevant distributions are shown in Fig.~\ref{fig:2DmassPTsq-colz}.
Overall, there are 566\,095 events in the signal sample and 56\,654 in the control sample.

The signal yields are determined in each rapidity interval by 
a two-dimensional unbinned extended maximum-likelihood
 fit~\cite{Verkerke:2003ir,James:1975dr} in mass and \ptsq.
The fit model
comprises the signal \jpsi and \psitwos components;
the continuum QED background;
the \psitwos and \chic feed-downs; 
and the inelastic PD background.
Prior to carrying out the fit in each rapidity interval, fits to the whole signal and control data samples
(referred to as the full sample) are performed
and their result\added{s} are used to constrain nuisance parameters that cannot be determined accurately in
low-yield rapidity regions, as described below.

The CEP and PD \jpsi and \psitwos mass peaks are each modelled with a Gaussian function,
modified to have power-law tails on both sides~\cite{Skwarnicki:1986xj}.
The difference in the means of the two Gaussian components is fixed according
to the known
mass difference of the two resonances~\cite{PDG2024}.
Their widths are constrained to scale linearly with the energy release in the respective decays~\cite{LHCb-PAPER-2021-008}.
The tail parameters are shared between the two peaks and Gaussian-constrained to the values
determined in the fits to the full sample.

The CEP \ptsq shape is independent of the mass and is described by an exponential function,
as expected by Regge theory~\cite{Collins:1977jy}
and measured in previous experiments, notably at HERA~\cite{Spiesberger:2018vki}.
The slopes of the \jpsi and \psitwos exponentials are left free to float in each rapidity interval.

The \ptsq distribution of the PD \jpsi and \psitwos mesons is modelled with a power-law function proportional to $(1+(b_\text{pd}/n_\text{pd})\ptsq)^{-n_\text{pd}}$, as measured by
the H1 experiment~\cite{H1:2013okq}. 
This function follows approximately an exponential of slope $-b_\text{pd}$, modified by the empirical parameter $n_\text{pd}$.
Alternative models are discussed in Sec.~\ref{sec:Syst}.
In addition, the PD contribution contains a nonresonant component which is modelled by an
exponential shape in mass and the above-mentioned power-law model for \ptsq.
The parameters of the three power-laws are different for the \jpsi, \psitwos, and nonresonant dimuon components.

\begin{figure}[t]
	\centering
        \includegraphics[width=0.48\textwidth]{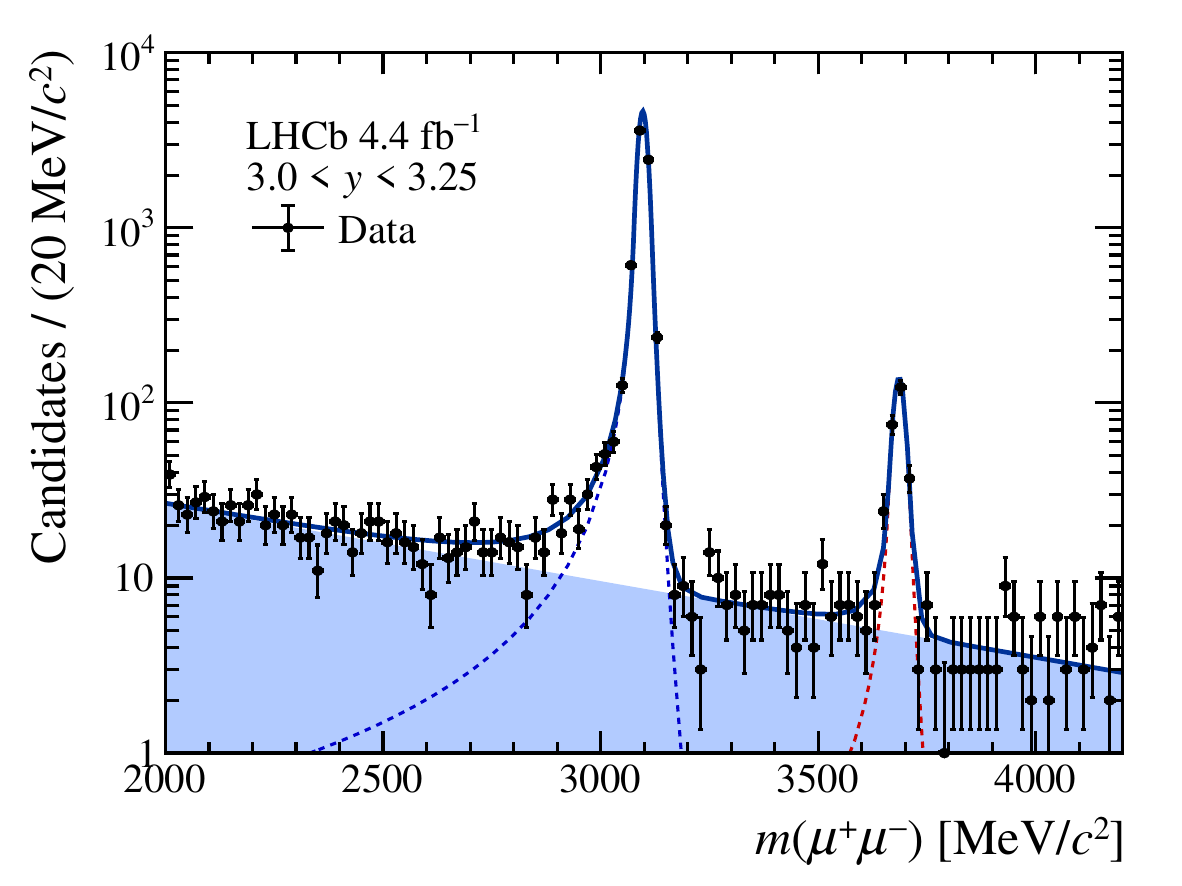}\hskip 0.03\textwidth%
        \includegraphics[width=0.48\textwidth]{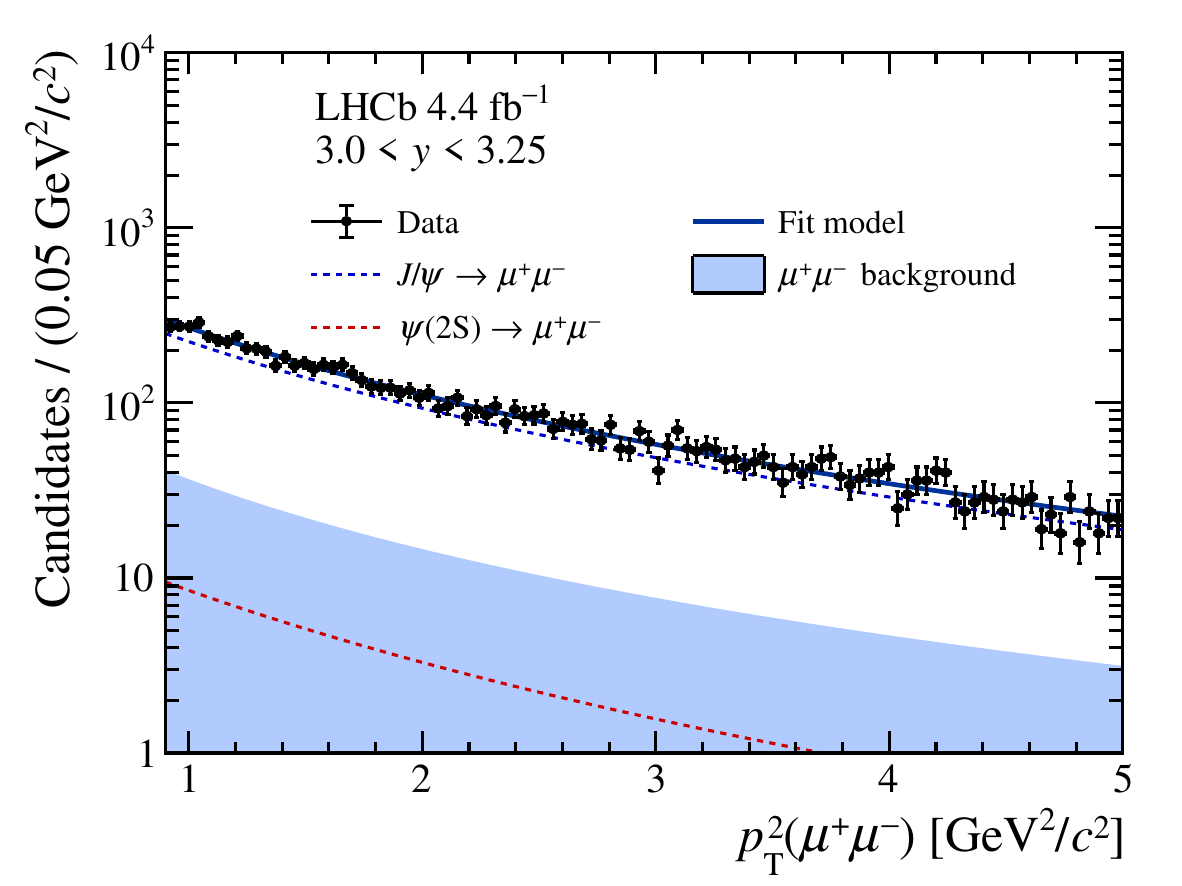}
	\caption{Distributions of (left) mass and (right) $p_{\mathrm{T}}^2$  of data in the control sample for rapidity interval $3.0<y<3.25$. The fit described in the text is superimposed.}
	\label{fig:PD_3_0}
\end{figure}

The parameters of the PD components are first determined by a fit to the control sample in each rapidity interval;
an example fit is shown in Fig.~\ref{fig:PD_3_0}.
All parameters are free to vary in these fits except for the signal tail parameters, as explained above.

The PD models are then used as input in the fits to the signal sample.
The \ptsq shapes of the PD components are fixed to the
values obtained on the corresponding control sample, while 
the PD \jpsi and \psitwos mass shapes are forced to be identical to those
of the CEP signals.
The relative fractions of the two charmonia and the dimuon background are constrained from the fit to the control sample.

Exclusive continuum, or nonresonant dimuon production, is a QED process that takes place via the fusion of two photons.
The dimuon pair produced in this form has low dimuon mass and a \ptsq shape sharply peaked towards zero.
The mass and \ptsq distributions are correlated and therefore a two-dimensional histogram, which is obtained from simulation
and validated with low-\ptsq data, is used in the fit.

The \jpsi yield is affected by feed-down from higher-mass charmonium states, which is accounted for in the fit.
The feed-down from \decay{\psitwos}{\jpsi X} decays is partially
suppressed by the VELO and SPD multiplicity requirements.
The yield of the remaining feed-down is determined from simulation of inclusive \decay{\psitwos}{\jpsi X} processes,
normalised by the \psitwos yield measured in each rapidity interval.
Bin migration is taken into account via a migration matrix determined from simulation.
An iterative procedure is applied to first determine the rapidity-dependent \psitwos yield and then
its contribution to the \jpsi yield. 
In practice, two steps are sufficient for the convergence of the procedure. 

\begin{figure}[b]
  \def\ww{0.48}
  \centering
    \includegraphics[width=\ww\textwidth]{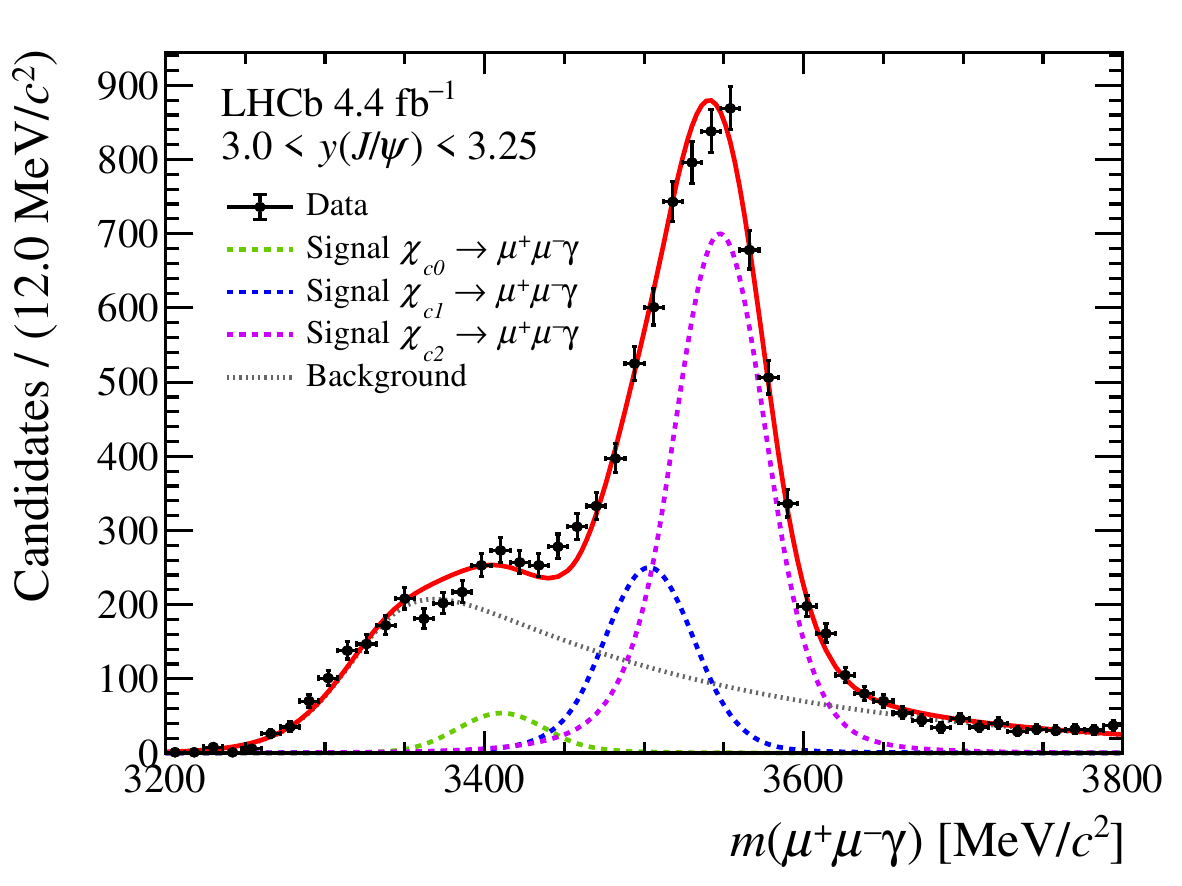}\hskip 0.03\textwidth%
    \includegraphics[width=\ww\textwidth]{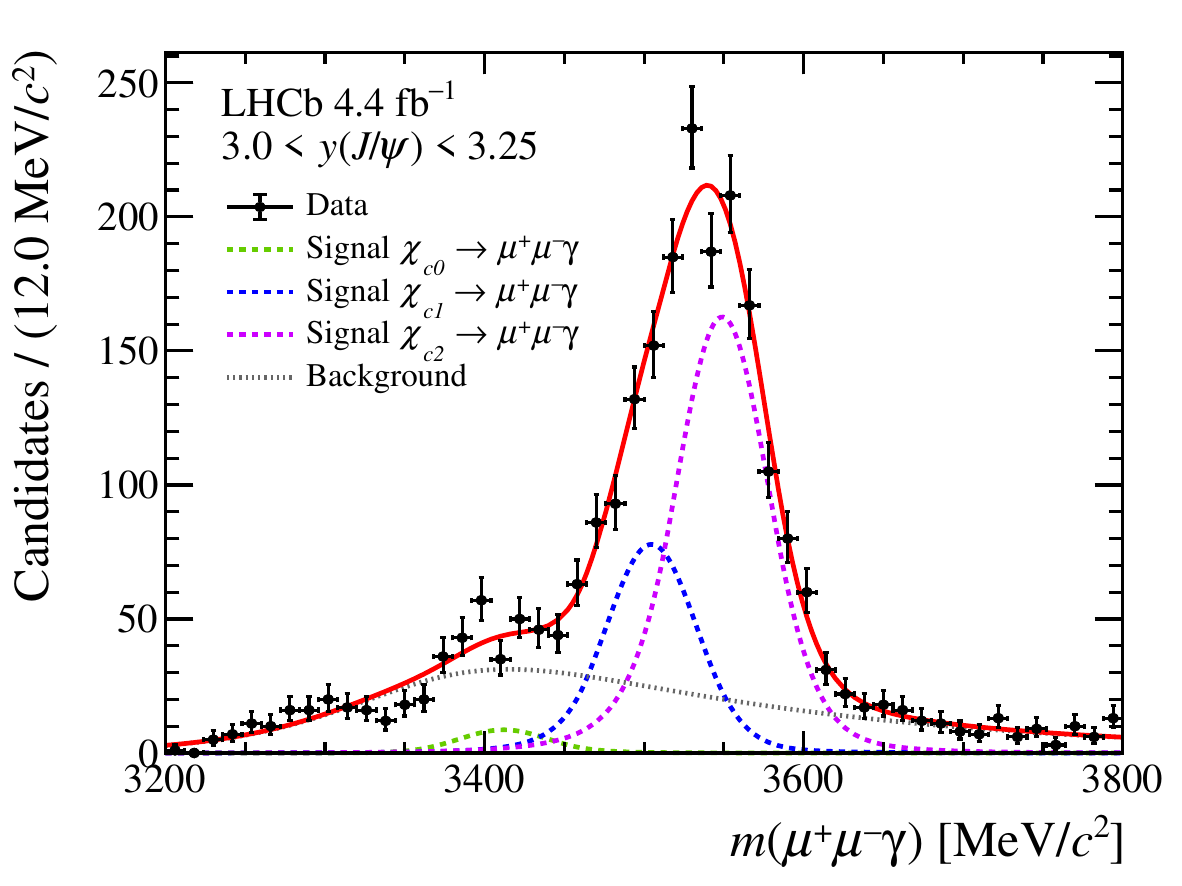}%
  \caption{Fit to the $J\mskip -3mu/\mskip -2mu\psi\gamma$ mass distribution with the $J\mskip -3mu/\mskip -2mu\psi$ meson in $3.0<y<3.25$ for (left) signal and (right) control samples.}
  \label{fig:Chic_3.0}
\end{figure}
The normalisation of the
feed-down from \decay{\chic}{\jpsi\gamma} decays is determined by reconstructing $\jpsi\gamma$ candidates in data. The same \jpsi selection as for CEP and PD candidates is used, except that the veto on additional photons is removed.
Instead, photons with transverse energy in excess of 75\mev are combined with \jpsi candidates to form  \chic candidates.
In each $y$ interval, where $y$ is the rapidity of the \jpsi meson, and separately for the signal and control samples, 
the \chicJ ($J=0,1,2$) yields are determined from a fit to the resulting
mass distribution. 
Fits to the \chic samples are shown in Fig.~\ref{fig:Chic_3.0}.
The three \chicJ ($J=0,1,2$) mass peaks are each modelled with a Crystal Ball function~\cite{Skwarnicki:1986xj}
with the tail parameters fixed from simulation.
The peak of the Gaussian is free in the fit to account for imperfect photon energy calibration, but the shift
with respect to the known masses of the \chic mesons~\cite{PDG2024}
is constrained to be the same for all three states.
The shift varies between 6 and 10\mevcc \added{(with typical statistical uncertainties between $0.5$ and $1\mevcc$)} depending on the rapidity interval.
The background is a mixture of partially reconstructed \psitwos decays, such as \decay{\psitwos}{\jpsi\piz\piz}, and
random combinations of \jpsi mesons and calorimeter clusters.
The same empirical function as in Ref.~\cite{LHCb-PAPER-2023-028} is used
as a model for the sum of these contributions.

The contribution from \chiczero mesons is small, while those of \chicone and \chictwo mesons dominate.
The empirical background model does not describe perfectly the mass distribution
in the region below 3400\mevcc,
which has a negligible effect on the total \chic yield.
Due to the limited photon energy resolution, the mass fit has little sensitivity to the
relative size of the \chicone and \chictwo contributions.
This ambiguity however does not affect  
the determination of the total \jpsi-from-$\chi_{c(1,2)}$ yield since
{\it (i)} the branching fractions of the \decay{\chi_{c(1,2)}}{\jpsi\Pgamma}
decays drop out in the ratio of \jpsi to $\chi_{c(1,2)}$ yields;
{\it (ii)} the
relative efficiencies for reconstructing \jpsi{}\Pgamma and \jpmm are equal for
\chicone and \chictwo,
as determined from simulation;
and {\it (iii)} the
\ptsq distributions of \jpsi from \chicone and \chictwo are found to be
equal in simulation and in data, which is checked by investigating the \ptsq
shape of candidates in the left and right halves of the $\chi_{c(1,2)}$ mass peak.
The \jpsi-from-$\chi_{c(1,2)}$ yield is therefore proportional to the
$\chi_{c(1,2)}$ yield in each rapidity interval.
This feed-down contribution is determined in the signal and control samples,
and the PD contribution is subtracted from that in CEP events
to determine the overall CEP \chic feed-down normalisation.

The \jpsi-from-\psitwos and \jpsi-from-\chic components are modelled in the CEP fit 
using the same mass model as for the \jpsi signal.
The \ptsq shapes are modelled with a single (double) exponential distribution for the \psitwos (\chic) feed-down,
which is determined from simulation that is validated by data.

\begin{figure}[tb]
	\centering
        \includegraphics[width=0.48\textwidth]{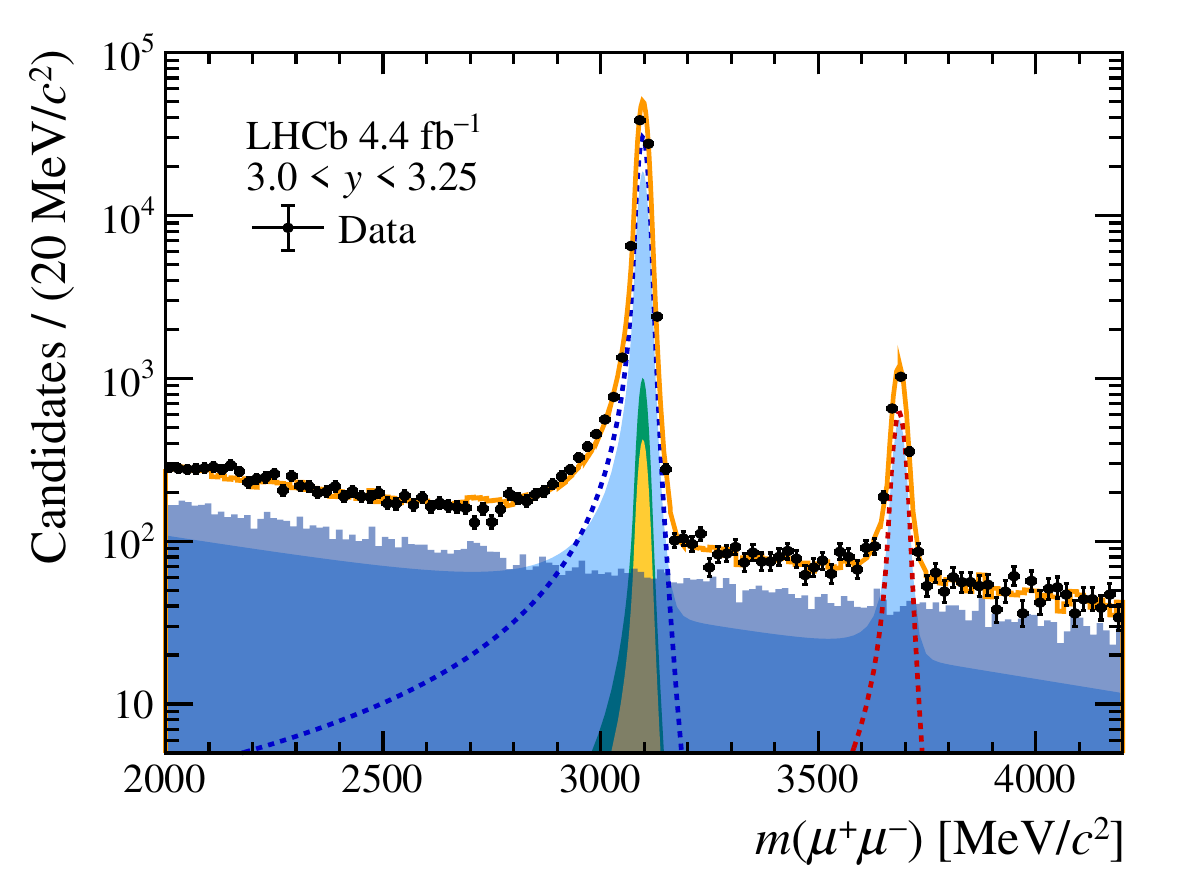}\hskip 0.03\textwidth%
        \includegraphics[width=0.48\textwidth]{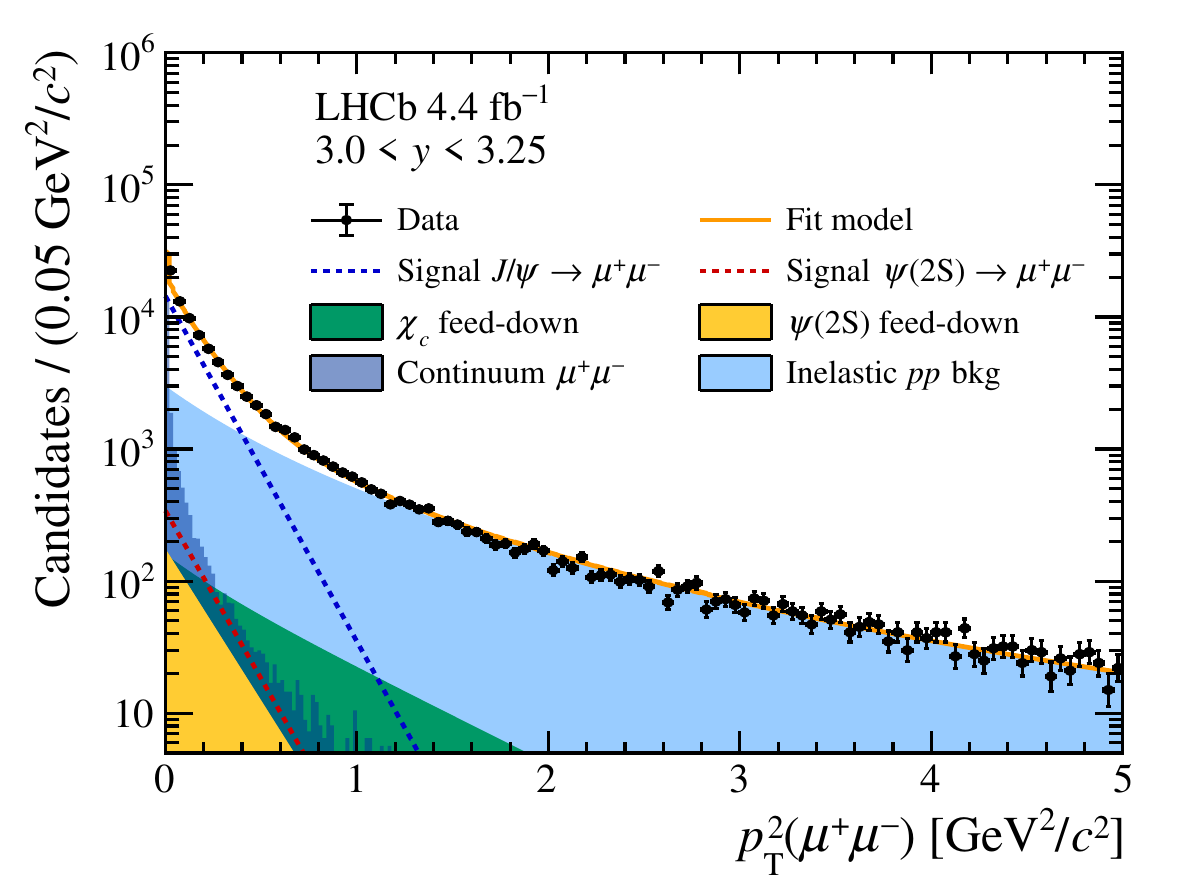}
	\caption{Distributions of (left) mass and (right) $p_{\mathrm{T}}^2$ of data in the signal sample for the rapidity interval $3.0<y<3.25$. The fit described in the text is superimposed.}
	\label{fig:CEP_3_0}
\end{figure}

The mass and \ptsq projections of the fit in the interval $3.0<y<3.25$ are shown in Fig.~\ref{fig:CEP_3_0}.
All intervals are shown in Fig.~\ref{fig:systUnc_ptsq_to_5_0-both_PDH1} in Appendix~\ref{App:fits}.
The parameters of interest are the CEP \jpsi and \psitwos yields, and the
slopes of their \ptsq shapes.
In total, $299\,100\pm2100$ \jpsi and $7420\pm130$ \psitwos elastically produced mesons are found
in the fit to the full rapidity range.

\section{Efficiencies}
\label{sec:Effs}
The signal yields are corrected for detection efficiencies using
simulation samples calibrated with data\added{~\cite{LHCb-DP-2013-002,LHCb-DP-2013-001,LHCb-DP-2014-001}}, except for the \HRC-related efficiencies,
which are estimated in data.

A tag-and-probe method, aimed at measuring single-muon efficiencies, is applied to account for the differences between simulation and data. 
The simulation sample is then weighted with the appropriate
correction factors~\cite{SanchezGras:2842971}.
In this method, a tag muon from the \jpsi candidate is required to pass all selection criteria, while the other muon
is used to measure the efficiency under investigation.
The same procedure is applied to calibration samples and simulation, and the latter is weighted by the ratio of those efficiencies.
The tracking, PID and hardware muon trigger efficiencies are calibrated in this manner.
As most efficiencies depend on muon kinematics, they are determined in regions of muon pseudorapidity
and transverse momentum, and separately for each year of data taking.
Depending on the considered $\pt,\eta$ region, correction factors range between 0.9 and 1.1 for tracking,
0.8 and 1.2 for PID, and 0.7 and 1.1 for muon trigger efficiencies, with uncertainties between 1\% and 3\%.

\begin{figure}[b]
  \centering
  \includegraphics[width=0.48\textwidth]{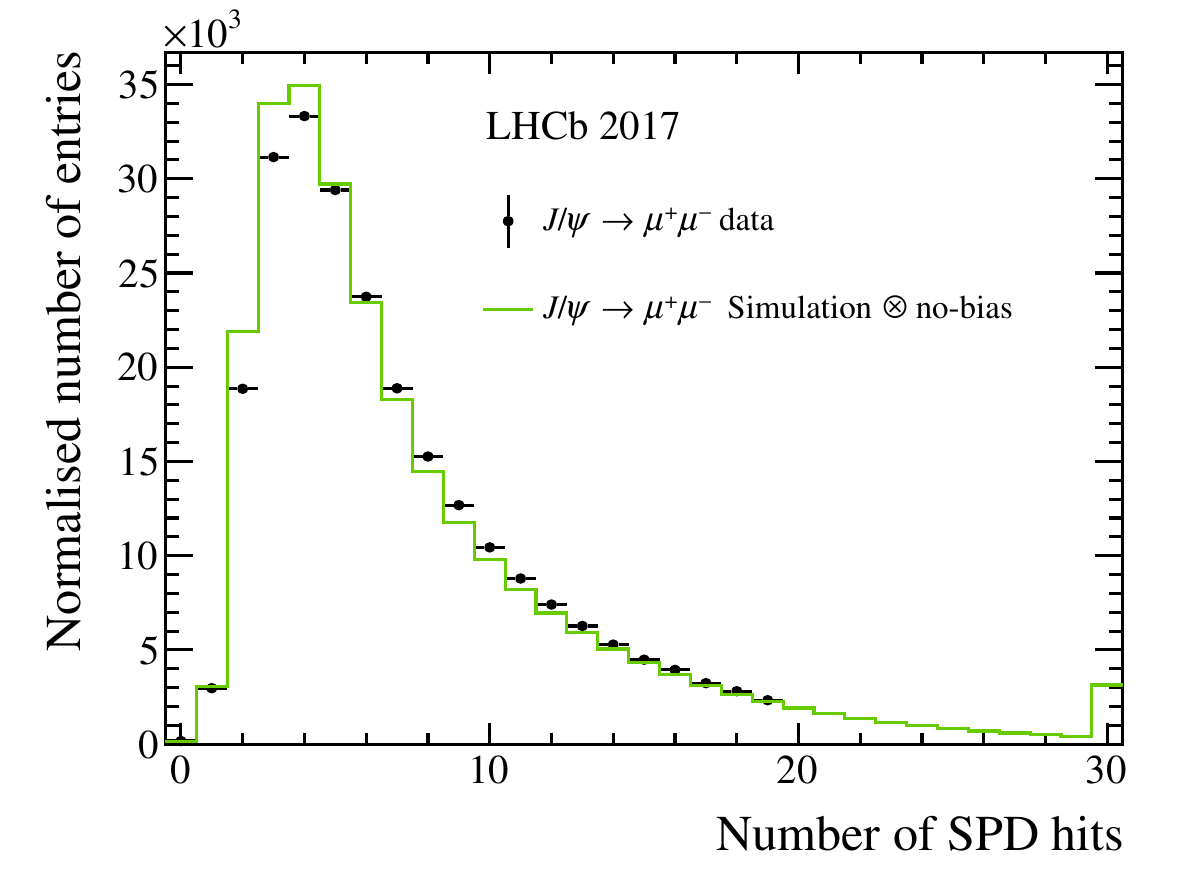}\hskip 0.03\textwidth%
  \includegraphics[width=0.48\textwidth]{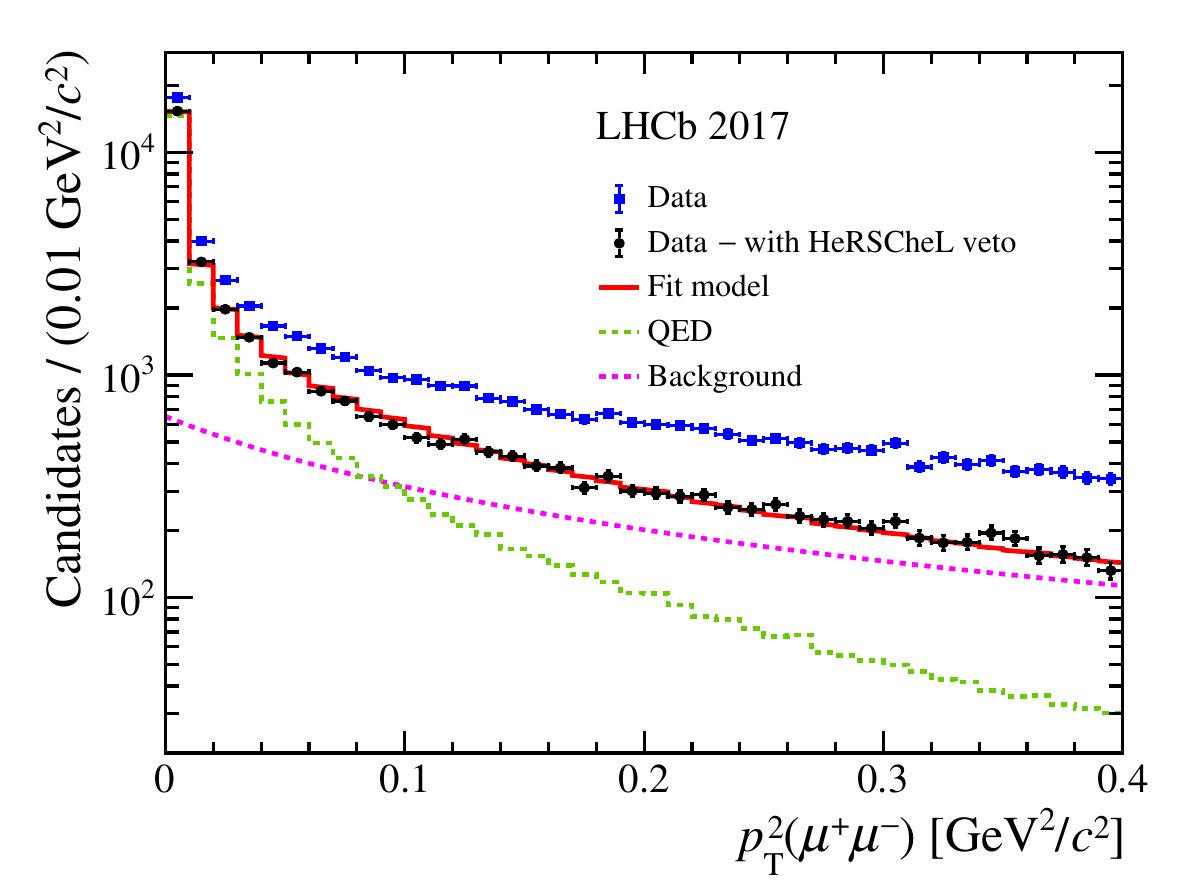}
  \caption{(Left) SPD multiplicity distributions in 2017 \jpmm data and modelling with simulation corrected using no-bias
    data from events with no beam crossing.
    The last bin contains the overflow events with 30 SPD hits or more.
    (Right) $p_{\mathrm{T}}^2$ distributions in 2017 data without and with the \textsc{HeRSCheL} requirement applied.
    The fit is only shown for the latter distribution.}
  \label{fig:SPDHits}\label{fig:Herschel}
\end{figure}

The hits in the SPD detector are due to charged particles reaching the detector, including those
produced by the preshower detector, and to spill-over from the previous $pp$ interaction.
In the case of CEP events, which have only two muon tracks, the latter component dominates;
however, it is not well modelled in simulation.
The SPD hit distribution due to spill-over is obtained from data events that were collected by random triggers
in unfilled bunch crossings that followed bunch crossings with a collision.
This sample is referred to as no-bias data
in the following.
The obtained distribution is convolved with the SPD multiplicity in \jpmm simulation and matches \added[reasonably]{sufficiently} well
the distribution observed in CEP data, especially the tail up to the cut value of 20 SPD hits, as shown in Fig.~\ref{fig:SPDHits}. 
\added{The effect of the remaining mismodelling is addressed in Sec~\ref{sec:Syst}.}
The fraction of events above this value defines the SPD trigger inefficiency, which is found to be independent of the dimuon kinematics.

The \herschel detector is not included in the simulation.
Its efficiency is determined using dimuon QED events, with and without the \herschel vetoes applied.
The \ptsq distributions are shown in Fig.~\ref{fig:Herschel}, emphasising the
fact that the \herschel requirement has little effect at \added[low]{vanishing} \ptsq, where QED backgrounds
dominate.
The efficiency is determined from the ratio of the QED components determined by the
fits to the distributions with and without the \herschel veto applied.
It is found to be between 85\% and 90\% depending on data-taking period.

Efficiencies for the requirement on the absence of additional VELO tracks or photons are taken from simulation and cross-checked
in data in the same way as for the \herschel veto efficiency.
They are close to unity.
The software trigger is fully efficient with respect to the offline selection.
The total efficiency varies between 40\% and 55\%, with the lowest values being
at the edges of the rapidity acceptance.

\section{Systematic uncertainties}
\label{sec:Syst}
Systematic uncertainties arise due to the size of the simulation and calibration samples,
and from the luminosity determination; they are mentioned in the sections above.
Systematic uncertainties related to modelling choices are described below.
All values are listed in Tables~\ref{tab:syst-both} and~\ref{tab:syst-b}.

\begin{table}[p]\centering
  \caption{Systematic and statistical uncertainties per rapidity interval for the $J\mskip -3mu/\mskip -2mu\psi$ and \mbox{$\psi(2S)$} cross-section in percent. The luminosity uncertainty is listed separately. Values below 0.005\% are not shown.}\label{tab:syst-both}
  \resizebox{\textwidth}{!}{\input{tables/tableLaTeX_systematicTables_both}}
\end{table}
\begin{table}[t]
  \caption{Systematic and statistical uncertainties per rapidity interval on the  exponential slopes $b_{J\mskip -3mu/\mskip -2mu\psi}$ and $b_{\psi(2S)}$, in percent.
   Values below 0.005\% are not reported.}\label{tab:syst-b}
  \resizebox{\textwidth}{!}{\input{tables/makeSystematicTables_fit_both_bSlope}}
\end{table}
In the modelling of the number of SPD hits, $N_\text{SPD}$, the convolution of the no-bias data and the \jpmm simulation
is normalised to the distribution seen in data for $N_\text{SPD}<20$.
Alternatively, one could normalise it to get the best possible description of
the tail, \ie in the region $8\le N_\text{SPD}<20$.
The resulting change in SPD veto efficiency is reported as a systematic uncertainty.

The efficiencies of the \herschel, track, and photon vetoes, collectively called global event cuts (GEC) in Table~\ref{tab:syst-both}, are taken from the yield of QED events passing the veto.
The fit models the non-QED background with the power-law from Ref.~\cite{H1:2013okq},
but in this reduced \ptsq region the sum of two exponential functions also provides a good description.
The difference in measured efficiencies is taken as the associated systematic uncertainty.

The signal \jpsi and \psitwos mass peaks are described by a modified Gaussian function.
Other models, such as the sum of a Gaussian and a Crystal Ball function~\cite{Skwarnicki:1986xj}, are tried
but yield poor-quality fits, without however significantly affecting the signal yields.
No uncertainty is assigned. 
\added{The peak value of the \psitwos mass shape is constrained to be offset from the \jpsi peak by the known mass difference of the two mesons. A systematic uncertainty is estimated by scaling the offset proportionally to the energy release in each decay.}

\added[Similarly, the CEP signal]{The} \ptsq shape\added[ is]{s are} modelled by a single exponential, as expected from theory,
and there is no evidence of the need for another model.
The \ptsq shapes of the \jpsi from feed-down are modelled with exponential functions taken from simulation.
Alternatively, nonparametric distributions obtained from simulation are used, which yields
a small change in the feed-down contributions and thus the signal yields.
For the \chic feed-down, a determination of the contribution from each $\chic$ state would be needed
if their \ptsq distributions were different.
The present data do not require this.
A dedicated CEP \chic study using converted photons would be needed to resolve the \chicone and \chictwo states.

The mass distribution of the inelastic $pp$ background is described by the same shape as for the signal \jpsi and \psitwos
plus an exponential to describe the nonresonant dimuon contribution.
A systematic uncertainty is estimated by changing the single exponential to the sum of two exponential functions.

Similarly, the \ptsq shape for each component of the inelastic $pp$ background is modelled with a power law\added{, as measured by the H1 experiment. However, the fit to the H1 dataset is not perfect at very low \ptsq~\cite{H1:2013okq}. Therefore other models are investigated}.
The sum of two exponentials \added[also provides]{is found to also provide} a reasonable  fit, though of slightly lower quality.
The systematic uncertainty due to this modelling is determined with pseudoexperiments.
In each rapidity interval the data are fit with the alternate mass-\ptsq model and then 
500 pseudodata samples are generated using the fit result as model.
These pseudosamples are then fit with the default model.
The resulting biases of the \jpsi and \psitwos yields are assigned as a systematic uncertainty.

Fit biases are tested in the same way.
This time the default shape is used to generate 500 pseudodata samples
which are fit with the same model.
The variation of yields is compatible
with the uncertainty returned from the fit, and the
biases are negligible in comparison.
No uncertainty is assigned.

Table~\ref{tab:syst-both} shows the values of all the systematic uncertainties previously discussed
per rapidity interval for the \jpmm and \psimm cross-section measurements.
All uncertainties are assumed to be uncorrelated between rapidity intervals,
except those for the SPD and \HRC multiplicities, and for the luminosity, which are fully correlated.

The slope of the signal \ptsq shape is only affected by changes in the fit model,
leading to the systematic uncertainties listed in Table~\ref{tab:syst-b}.

\section{Results}
\label{sec:Results}
The signal yields determined by the two-dimensional fit, the efficiencies and the resulting
differential  \decay{pp}{p\jpsi p} and \decay{pp}{p\psitwos p} cross-sections are reported
in Tables~\ref{tab:crossSection-theory-Jpsi2mumu} and~\ref{tab:crossSection-theory-Psi2S2mumu} in Appendix~\ref{App:numbers}.
Summing over all rapidity intervals, the total integrated cross-sections for charmonia with $2.0<y<4.5$
and muons with $2.0<\eta<4.5$ are 
\begin{align*}
  \input{tables/total_Jpsi2mumu} \\
  \input{tables/total_psi2S2mumu} 
\end{align*}
where the first uncertainties are statistical, the second systematic and the third are due to the luminosity determination.
These values are more precise than those reported in the previous analysis of exclusive \jpsi and \psitwos production at \mbox{$\sqs = 13 \tev$}~\cite{LHCb-PAPER-2018-011} and are compatible \added{with the previous results} at the level of $1.5\,\sigma$.

\begin{figure}[t]
	\begin{center}
		\begin{tabular}{c}
		  \includegraphics[width=0.48\textwidth]{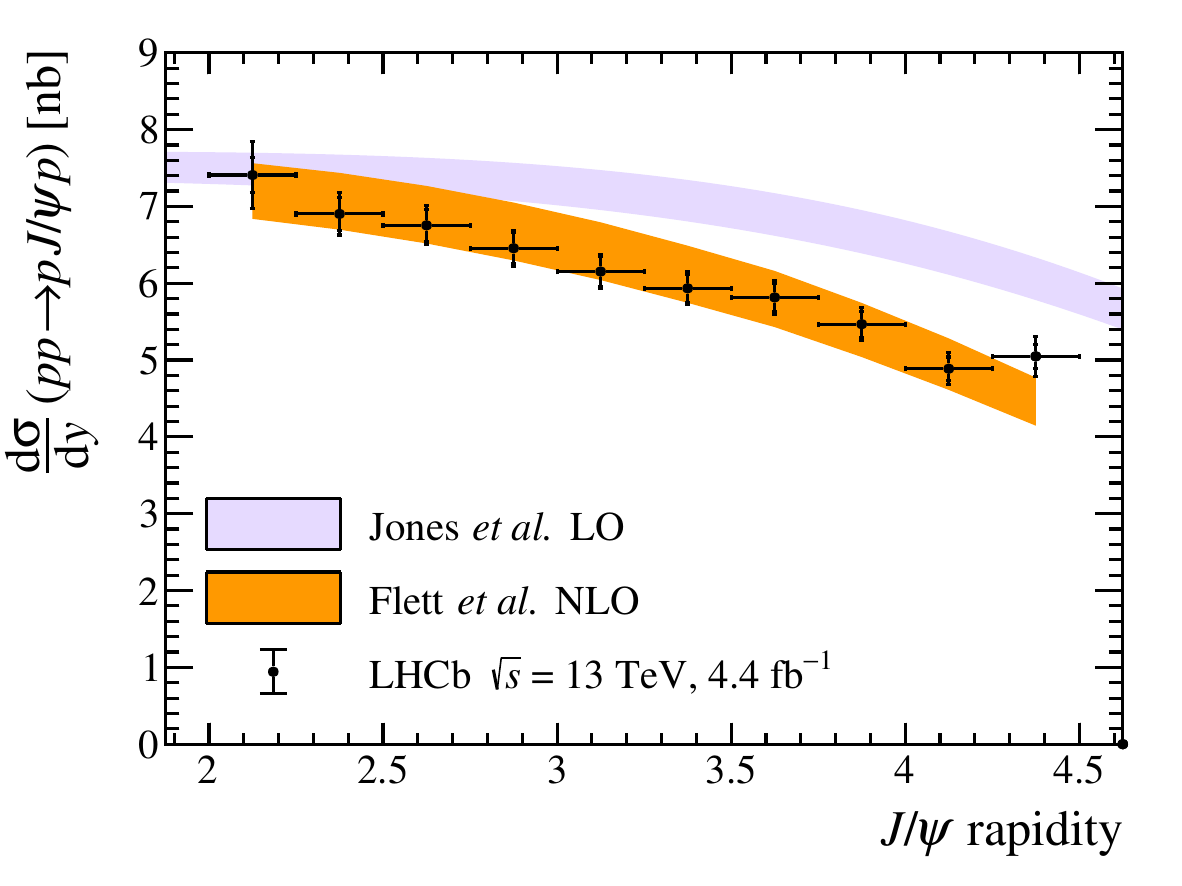}
                  \includegraphics[width=0.48\textwidth]{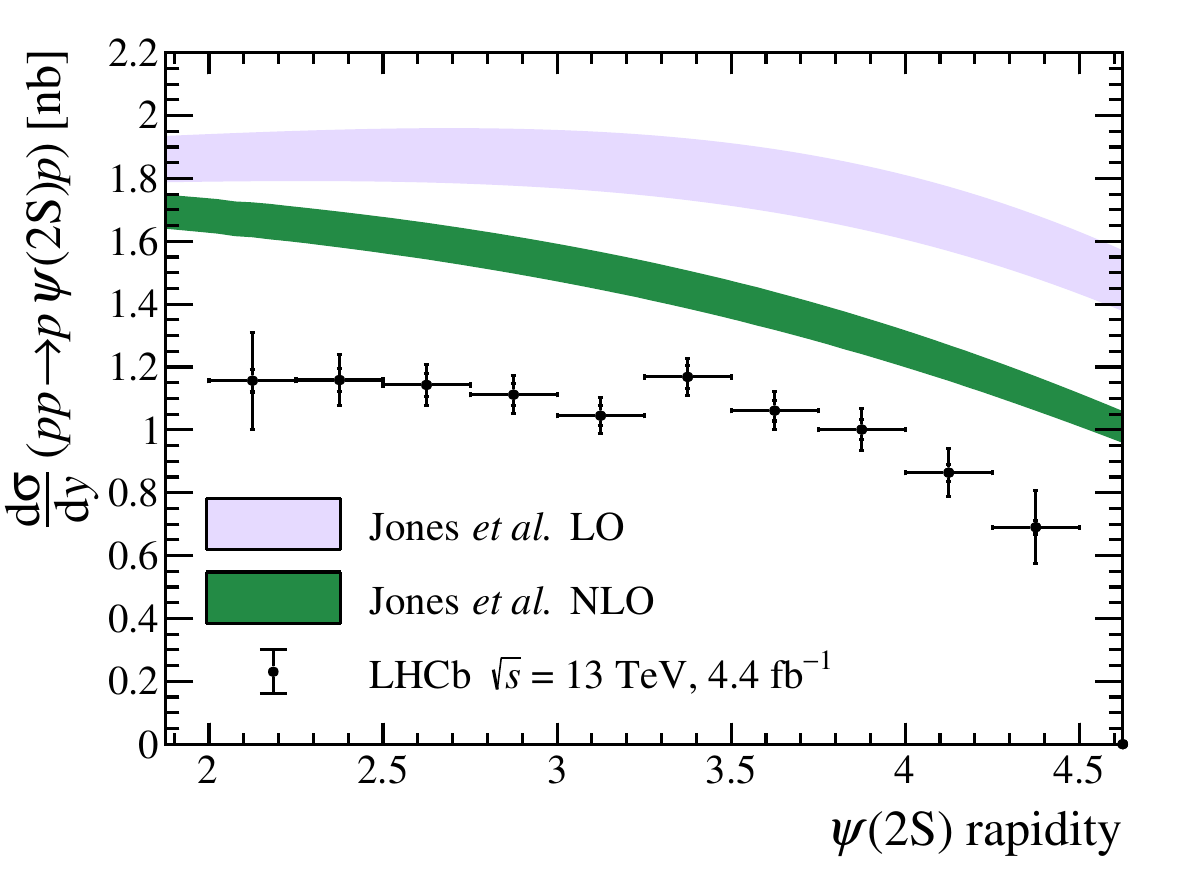}
			
		\end{tabular}
		\caption{Differential cross-section for (left) $J\mskip -3mu/\mskip -2mu\psi$ and (right) $\psi(2S)$ mesons.
Theoretical predictions from Jones \etal~\cite{Jones:2013pga,Jones:2013eda} and Flett \etal~\cite{Flett:2020duk} are shown for comparison.} 
		\label{fig:crossSectionMM}
	\end{center}
\end{figure}

The measured cross-sections are corrected for the \jpmm and \psiee branching fractions~\cite{PDG2024} and the detector acceptance of the two muons, using simulation.
The branching fraction for  \psiee is used under the assumption of lepton universality as it is more
precise than that of \psimm. 
The resulting differential cross-sections are shown in Fig.~\ref{fig:crossSectionMM}.
Theoretical predictions are shown for comparison.
\added[While the]{The} 
\jpsi cross-section agrees with the NLO prediction~\cite{Flett:2020duk}, 
\added[which
improves on the LO prediction]{which is posterior to the previous 13\tev measurement~\cite{LHCb-PAPER-2018-011}}. 
\added{On the other hand, the} \psitwos cross-section is significantly lower than both the LO and NLO predictions~\cite{Jones:2013eda}\added{, which predate the LHCb measurement. The present result calls for an updated calculation of the differential \psitwos prediction}.

The \jpsi cross-sections are used to determine the photoproduction cross-section as a function of the photon-proton energy,
which is reported in Appendix~\ref{App:photo}.

\begin{figure}[tb]
	\centering
	\includegraphics[width=0.48\textwidth]{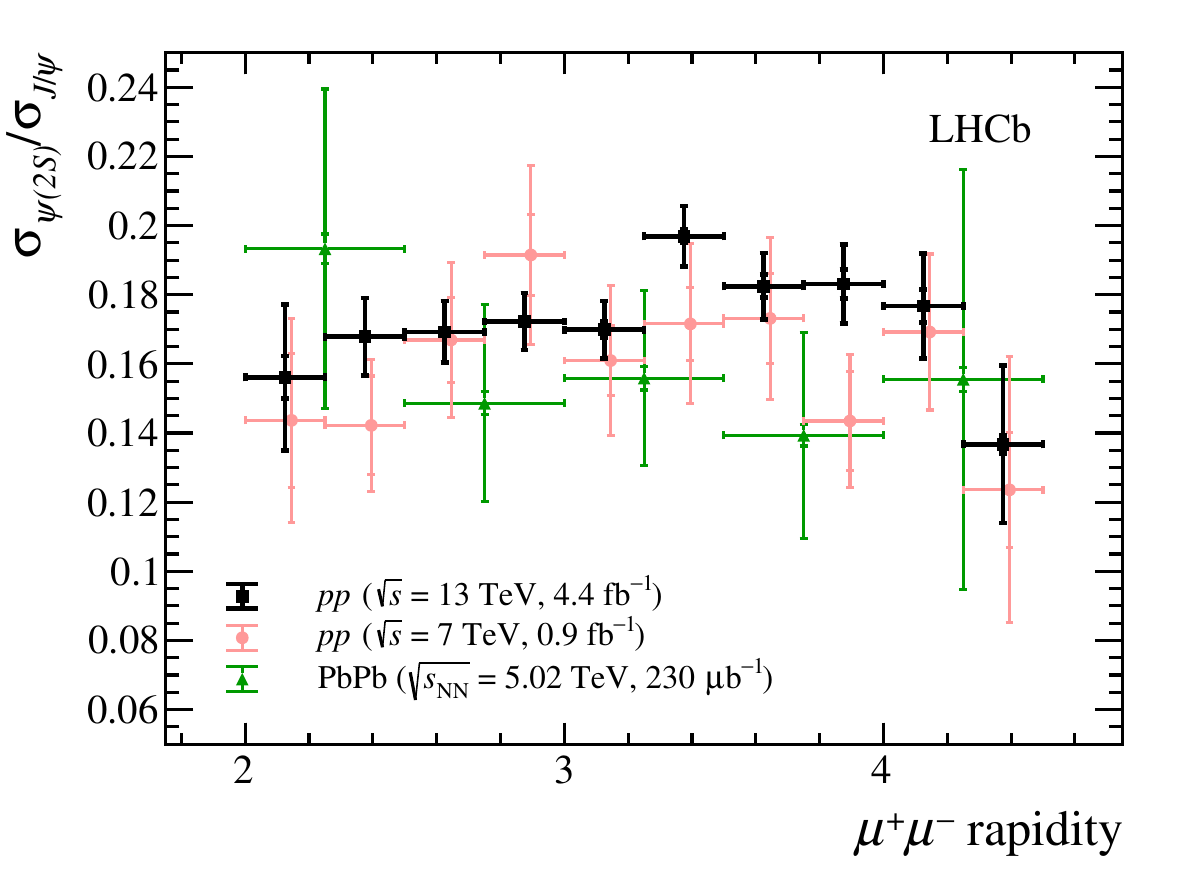} 
	\caption{Measured ratio of \psitwos and $J\mskip -3mu/\mskip -2mu\psi$ cross-sections per rapidity interval.
          The results from PbPb collisions at $\sqrt{s_{\scriptscriptstyle\text{NN}}}=5.02\,\text{Te\kern -0.1em V}$~\cite{LHCb-PAPER-2022-012}
          and $pp$ collisions at $\sqrt{s}=7\,\text{Te\kern -0.1em V}$~\cite{LHCb-PAPER-2013-059} are shown for comparison.
        The latter data points are slightly offset horizontally to increase visibility. } 
	\label{fig:ratioMM}
\end{figure}

The ratio of \psitwos and \jpsi cross-sections integrated over rapidity is found to be
\input{tables/ratio}%
where the last uncertainty is due to the knowledge of the branching fractions.
The luminosity uncertainty cancels in the ratio.
The ratio is shown in rapidity intervals in Fig.~\ref{fig:ratioMM} and
agrees with measurements by the LHCb collaboration in
$pp$ collisions at $\sqs=7\tev$~\cite{LHCb-PAPER-2013-059} and
PbPb collisions at $\sqsnn=5.02\tev$~\cite{LHCb-PAPER-2022-012}.
Its average is consistent with those measured by the H1 and ZEUS collaborations~\cite{H1:2002yab,ZEUS:2016awm}.

\begin{figure}[t]\centering
  \includegraphics[width=0.48\textwidth]{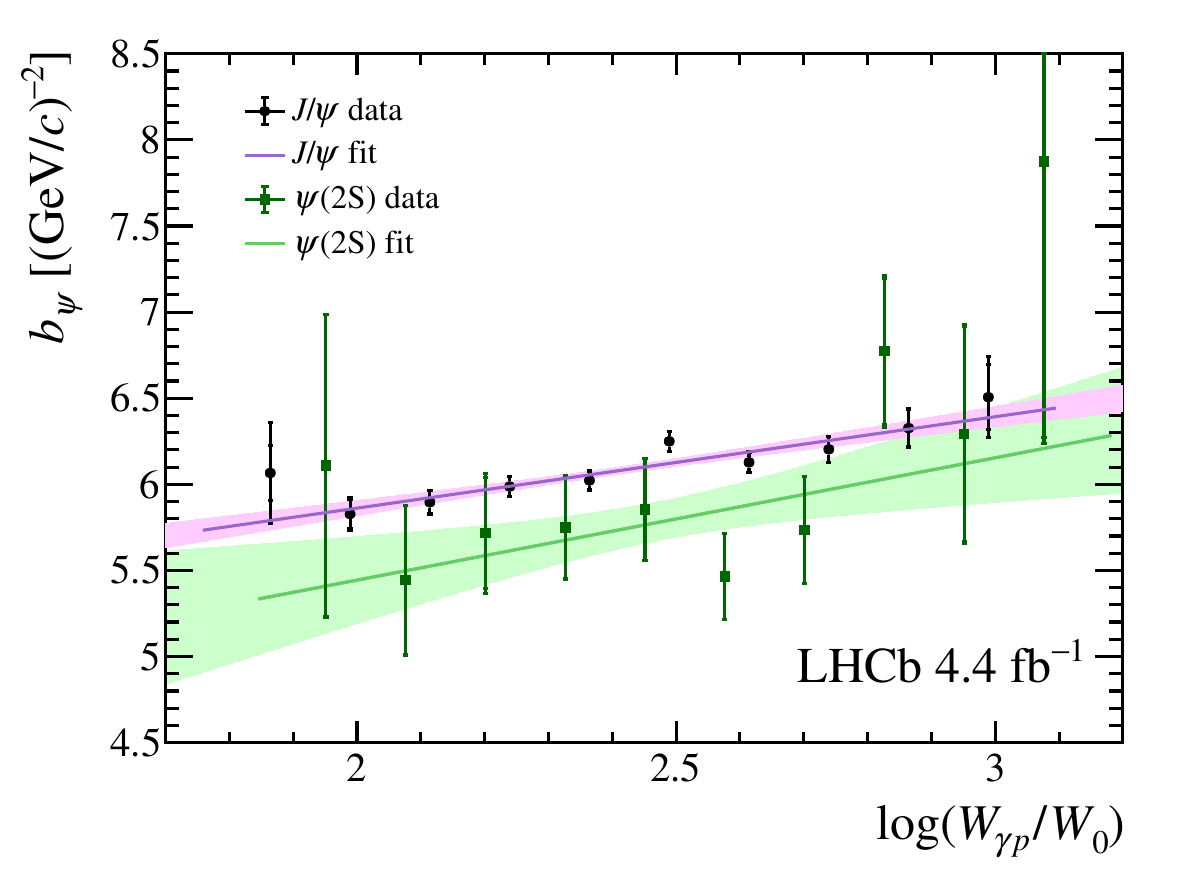}		
  \includegraphics[width=0.48\textwidth]{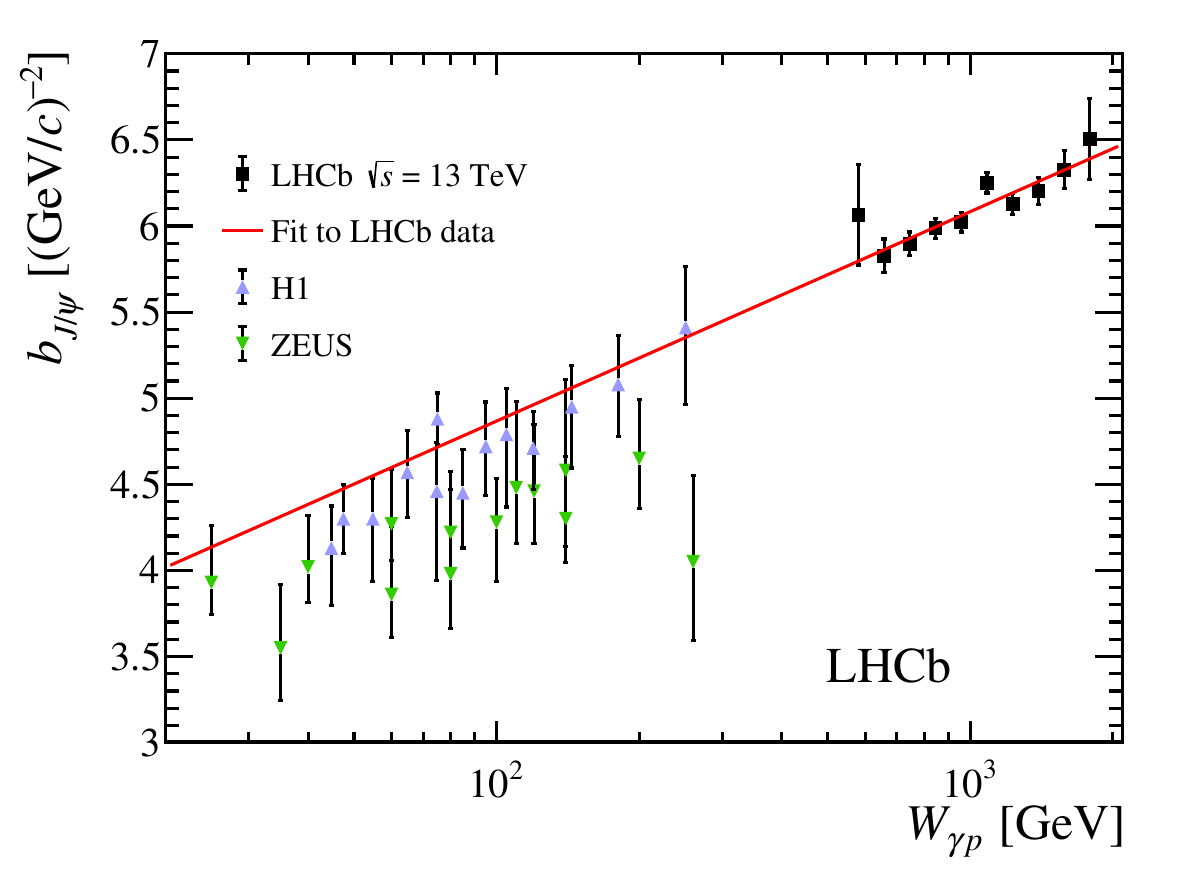}		
  \caption{(Left) linear fit to the logarithmic dependence of $b_{J\mskip -3mu/\mskip -2mu\psi}$ and $b_{\psi(2S)}$ with respect
    to the photon-proton
    energy $W_{\gamma p}$ for $J\mskip -3mu/\mskip -2mu\psi$ and \psitwos production.
    The shaded areas represent the 68\% C.L. fit uncertainties. 
    (Right) measured $b$ slopes for $J\mskip -3mu/\mskip -2mu\psi$ production by the LHCb (this paper), H1~\cite{H1:2005dtp,H1:2013okq} and ZEUS~\cite{ZEUS:2002wfj} experiments.
          Superimposed is a line with the slope resulting from the fit to the LHCb data.} 
  \label{fig:shrinkageFit}
  \label{fig:shrinkageB}
\end{figure}

The photoproduction cross-section has an exponential behaviour versus \ptsq:
\mbox{${\rm d}\sigma/{\rm d}\ptsq \sim e^{-b\ptsq}$}.
The exponential slope $b$ can be parameterised as
\begin{equation}\label{eq:slopeb-2}
	\begin{aligned}
		b = b_0 + 4 \alpha^{\prime} \log \left(\frac{W_{\gamma p}}{W_0} \right)\,,
	\end{aligned}
\end{equation}
where, in Regge theory, $\alpha^{\prime}$ is the slope of the pomeron trajectory,
$W_{\gamma p}$ is the photon-proton centre-of-mass energy defined in Appendix~\ref{App:photo},
$W_0$ is typically taken to be $W_0 = 90 \gev$,
and $b_0$ is determined experimentally.

A linear fit to the $b$ slopes in intervals of rapidity is shown in Fig.~\ref{fig:shrinkageB}.
The intercepts and slopes are determined to be
\input{tables/pomeron}%
where the first uncertainty is statistical and the second systematic. The systematic uncertainties are obtained by taking the difference of the central value when the fit is performed with and without the systematic uncertainties accounted for in the fit.
The fit to \jpsi data agrees with previous determinations in $ep$ collisions~\cite{H1:2005dtp,H1:2013okq,ZEUS:2002wfj,H1:2020lzc} but is below the prediction of Ref.~\cite{Donnachie:1999qv}.

\section{Conclusion}\label{sec:Conclusion}

This paper presents the first measurement of the exclusive \psitwos cross-section in $pp$ collisions at $\sqs=13\tev$ in ten intervals of rapidity between $2.0$ and $4.5$.
The corresponding \jpsi cross-section is updated and the rapidity-dependent ratio \added{of \psitwos and \jpsi production} is determined for the first time.
The results are consistent but more precise than those of Ref.~\cite{LHCb-PAPER-2018-011}. 
When expressed as a function of the photon-proton energy, the cross-sections are found to be consistent with previous measurements,
but the \psitwos cross-section is below theory predictions.

For the first time, the dependence of the \jpsi and \psitwos cross-sections on $\ptsq\sim \Delta t$, where $\Delta t$ is the total transverse momentum transfer, is determined in $pp$ collisions
and is found consistent with \added{and more precise than} the behaviour observed at HERA~\cite{H1:2005dtp,H1:2013okq,ZEUS:2002wfj,H1:2020lzc}.

%% file: tables/tableLaTeX_systematicTables_both.tex
\begin{tabular}{lc@{\ }cc@{\ }cc@{\ }cc@{\ }cc@{\ }c} 
\toprule 
\multicolumn{1}{r}{\boldmath \bf $y$ bin} & \multicolumn{2}{c}{\bf 2.0--2.25} & \multicolumn{2}{c}{\bf 2.25--2.5} & \multicolumn{2}{c}{\bf 2.5--2.75} & \multicolumn{2}{c}{\bf 2.75--3.0} & \multicolumn{2}{c}{\bf 3.0--3.25} \\ 
{\bf Source / state} & \jpsi & \psitwos & \jpsi & \psitwos   & \jpsi & \psitwos   & \jpsi & \psitwos   & \jpsi & \psitwos \\  
\midrule 
\multicolumn{11}{c}{\bf Uncorrelated uncertainties}\\ 
\midrule 
{\boldmath \bf Simulation sample size} &\phz 0.35 &\phz 0.6\phz &\phz 0.16 &\phz 0.28 &\phz 0.12 &\phz 0.21 &\phz 0.10 &\phz 0.18 &\phz 0.10 &\phz 0.18\\ 
{\boldmath \bf Bin migration} &\phz 0.03 &\phz 0.08 &  &\phz 0.01 &  &  &\phz 0.02 &\phz 0.05 &\phz 0.04 &\phz 0.05\\ 
{\boldmath \bf Muon efficiency} &\phz 2.0\phz &\phz 1.9\phz &\phz 1.7\phz &\phz 1.6\phz &\phz 1.6\phz &\phz 1.5\phz &\phz 1.6\phz &\phz 1.5\phz &\phz 1.5\phz &\phz 1.5\phz\\ 
{\boldmath \bf Mass: PD shape} &\phz 0.06 &\phz 0.22 &\phz 0.01 &\phz 0.17 &\phz 0.10 &\phz 0.22 &\phz 0.10 &\phz 0.18 &\phz 0.19 &\phz 0.33\\ 
{\boldmath \bf Mass: \psitwos offset} &\phz 0.03 &\phz 0.17 &\phz 0.02 &\phz 0.15 &\phz 0.02 &\phz 0.02 &  &  &  &\phz 0.01\\ 
{\boldmath \bf \ptsq: \chic feed-down} &\phz 0.01 &\phz 0.01 &\phz 0.01 &\phz 0.06 &\phz 0.03 &\phz 0.01 &\phz 0.04 &\phz 0.01 &\phz 0.02 &\phz 0.03\\ 
{\boldmath \bf \ptsq: \psitwos feed-down} &\phz 0.12 &\phz 0.04 &\phz 0.01 &\phz 0.09 &\phz 0.01 &\phz 0.02 &\phz 0.02 &\phz 0.02 &\phz 0.03 &\phz 0.04\\ 
{\boldmath \bf \ptsq: PD shape} &\phz 3.5\phz &\phz 1.8\phz &\phz 0.20 &\phz 0.4\phz &\phz 0.30 &\phz 0.11 &\phz 0.01 &\phz 0.7\phz &\phz 0.32 &\phz 1.2\phz\\ 
\midrule 
{\boldmath \bf Total uncorrelated} &\phz 4.0\phz &\phz 2.7\phz &\phz 1.8\phz &\phz 1.7\phz &\phz 1.7\phz &\phz 1.5\phz &\phz 1.6\phz &\phz 1.7\phz &\phz 1.5\phz &\phz 2.0\phz\\ 
\midrule 
\multicolumn{11}{c}{\bf Correlated uncertainties}\\ 
\midrule 
{\boldmath \bf GEC efficiency} &\phz 0.8\phz &\phz 0.8\phz &\phz 0.8\phz &\phz 0.8\phz &\phz 0.8\phz &\phz 0.8\phz &\phz 0.8\phz &\phz 0.8\phz &\phz 0.8\phz &\phz 0.8\phz\\ 
{\boldmath \bf GEC background} &\phz 0.8\phz &\phz 0.8\phz &\phz 0.8\phz &\phz 0.8\phz &\phz 0.8\phz &\phz 0.8\phz &\phz 0.8\phz &\phz 0.8\phz &\phz 0.8\phz &\phz 0.8\phz\\ 
{\boldmath \bf SPD hits efficiency} &\phz 0.03 &\phz 0.03 &\phz 0.03 &\phz 0.03 &\phz 0.03 &\phz 0.03 &\phz 0.03 &\phz 0.03 &\phz 0.03 &\phz 0.03\\ 
{\boldmath \bf SPD multiplicity shape} &\phz 0.15 &\phz 0.15 &\phz 0.15 &\phz 0.15 &\phz 0.15 &\phz 0.15 &\phz 0.15 &\phz 0.15 &\phz 0.15 &\phz 0.15\\ 
\midrule 
{\boldmath \bf Total correlated} &\phz 1.2\phz &\phz 1.2\phz &\phz 1.2\phz &\phz 1.2\phz &\phz 1.2\phz &\phz 1.2\phz &\phz 1.2\phz &\phz 1.2\phz &\phz 1.2\phz &\phz 1.2\phz\\ 
\midrule 
\midrule 
\multicolumn{11}{c}{\bf Total uncertainties}\\ 
\midrule 
{\boldmath \bf Systematic (excl. luminosity)} &\phz 4.2\phz &\phz 2.9\phz &\phz 2.1\phz &\phz 2.0\phz &\phz 2.0\phz &\phz 1.9\phz &\phz 1.9\phz &\phz 2.0\phz &\phz 1.9\phz &\phz 2.3\phz\\ 
{\boldmath \bf Luminosity} &\phz 2.9\phz &\phz 2.9\phz &\phz 2.9\phz &\phz 2.9\phz &\phz 2.9\phz &\phz 2.9\phz &\phz 2.9\phz &\phz 2.9\phz &\phz 2.9\phz &\phz 2.9\phz\\ 
{\boldmath \bf Statistical} &\phz 2.3\phz & 12.6\phz &\phz 1.5\phz &\phz 6.1\phz &\phz 1.1\phz &\phz 4.6\phz &\phz 1.0\phz &\phz 4.1\phz &\phz 0.9\phz &\phz 4.0\phz\\ 
\bottomrule 
\\ 
\toprule 
\multicolumn{1}{r}{\boldmath \bf $y$ bin} & \multicolumn{2}{c}{\bf 3.25--3.5} & \multicolumn{2}{c}{\bf 3.5--3.75} & \multicolumn{2}{c}{\bf 3.75--4.0} & \multicolumn{2}{c}{\bf 4.0--4.25} & \multicolumn{2}{c}{\bf 4.25--4.5} \\ 
{\bf Source / state} & \jpsi & \psitwos & \jpsi & \psitwos   & \jpsi & \psitwos   & \jpsi & \psitwos   & \jpsi & \psitwos \\  
\midrule 
\multicolumn{11}{c}{\bf Uncorrelated uncertainties}\\ 
\midrule 
{\boldmath \bf Simulation sample size} &\phz 0.09 &\phz 0.18 &\phz 0.10 &\phz 0.20 &\phz 0.12 &\phz 0.25 &\phz 0.16 &\phz 0.35 &\phz 0.32 &\phz 0.7\phz\\ 
{\boldmath \bf Bin migration} &\phz 0.01 &\phz 0.02 &\phz 0.02 &\phz 0.02 &  &\phz 0.03 &\phz 0.03 &\phz 0.10 &\phz 0.09 &\phz 0.14\\ 
{\boldmath \bf Muon efficiency} &\phz 1.4\phz &\phz 1.5\phz &\phz 1.5\phz &\phz 1.5\phz &\phz 1.6\phz &\phz 1.5\phz &\phz 1.6\phz &\phz 1.7\phz &\phz 1.8\phz &\phz 2.1\phz\\ 
{\boldmath \bf Mass: PD shape} &\phz 0.17 &\phz 0.28 &\phz 0.13 &\phz 0.15 &\phz 0.13 &\phz 0.32 &\phz 0.06 &\phz 0.20 &\phz 0.03 &\phz 0.10\\ 
{\boldmath \bf Mass: \psitwos offset} &\phz 0.03 &  &\phz 0.03 &\phz 0.02 &\phz 0.02 &\phz 0.02 &\phz 0.06 &\phz 0.02 &\phz 0.04 &\phz 0.09\\ 
{\boldmath \bf \ptsq: \chic feed-down} &  &\phz 0.01 &\phz 0.02 &  &  &\phz 0.03 &\phz 0.03 &\phz 0.03 &\phz 0.03 &\phz 0.02\\ 
{\boldmath \bf \ptsq: \psitwos feed-down} &\phz 0.04 &\phz 0.01 &\phz 0.01 &\phz 0.02 &\phz 0.01 &\phz 0.04 &\phz 0.05 &\phz 0.01 &  &\phz 0.04\\ 
{\boldmath \bf \ptsq: PD shape} &\phz 0.6\phz &\phz 0.6\phz &\phz 0.8\phz &\phz 1.6\phz &\phz 1.0\phz &\phz 2.1\phz &\phz 1.5\phz &\phz 2.2\phz &\phz 1.5\phz &\phz 0.9\phz\\ 
\midrule 
{\boldmath \bf Total uncorrelated} &\phz 1.6\phz &\phz 1.6\phz &\phz 1.7\phz &\phz 2.2\phz &\phz 1.8\phz &\phz 2.6\phz &\phz 2.2\phz &\phz 2.8\phz &\phz 2.4\phz &\phz 2.4\phz\\ 
\midrule 
\multicolumn{11}{c}{\bf Correlated uncertainties}\\ 
\midrule 
{\boldmath \bf GEC efficiency} &\phz 0.8\phz &\phz 0.8\phz &\phz 0.8\phz &\phz 0.8\phz &\phz 0.8\phz &\phz 0.8\phz &\phz 0.8\phz &\phz 0.8\phz &\phz 0.8\phz &\phz 0.8\phz\\ 
{\boldmath \bf GEC background} &\phz 0.8\phz &\phz 0.8\phz &\phz 0.8\phz &\phz 0.8\phz &\phz 0.8\phz &\phz 0.8\phz &\phz 0.8\phz &\phz 0.8\phz &\phz 0.8\phz &\phz 0.8\phz\\ 
{\boldmath \bf SPD hits efficiency} &\phz 0.03 &\phz 0.03 &\phz 0.03 &\phz 0.03 &\phz 0.03 &\phz 0.03 &\phz 0.03 &\phz 0.03 &\phz 0.03 &\phz 0.03\\ 
{\boldmath \bf SPD multiplicity shape} &\phz 0.15 &\phz 0.15 &\phz 0.15 &\phz 0.15 &\phz 0.15 &\phz 0.15 &\phz 0.15 &\phz 0.15 &\phz 0.15 &\phz 0.15\\ 
\midrule 
{\boldmath \bf Total correlated} &\phz 1.2\phz &\phz 1.2\phz &\phz 1.2\phz &\phz 1.2\phz &\phz 1.2\phz &\phz 1.2\phz &\phz 1.2\phz &\phz 1.2\phz &\phz 1.2\phz &\phz 1.2\phz\\ 
\midrule 
\midrule 
\multicolumn{11}{c}{\bf Total uncertainties}\\ 
\midrule 
{\boldmath \bf Systematic (excl. luminosity)} &\phz 1.9\phz &\phz 2.0\phz &\phz 2.0\phz &\phz 2.5\phz &\phz 2.2\phz &\phz 2.8\phz &\phz 2.5\phz &\phz 3.0\phz &\phz 2.6\phz &\phz 2.6\phz\\ 
{\boldmath \bf Luminosity} &\phz 2.9\phz &\phz 2.9\phz &\phz 2.9\phz &\phz 2.9\phz &\phz 2.9\phz &\phz 2.9\phz &\phz 2.9\phz &\phz 2.9\phz &\phz 2.9\phz &\phz 2.9\phz\\ 
{\boldmath \bf Statistical} &\phz 0.9\phz &\phz 3.6\phz &\phz 1.0\phz &\phz 4.3\phz &\phz 1.2\phz &\phz 5.2\phz &\phz 1.6\phz &\phz 7.7\phz &\phz 2.5\phz & 16.2\phz\\ 
\bottomrule 
\end{tabular} 

%% file: tables/makeSystematicTables_fit_both_bSlope.tex
\begin{tabular}{lc@{\ }cc@{\ }cc@{\ }cc@{\ }cc@{\ }c} 
\toprule 
{\boldmath \bf $y$ bin} & \multicolumn{2}{c}{\bf 2.0--2.25} & \multicolumn{2}{c}{\bf 2.25--2.5} & \multicolumn{2}{c}{\bf 2.5--2.75} & \multicolumn{2}{c}{\bf 2.75--3.0} & \multicolumn{2}{c}{\bf 3.0--3.25} \\ 
{\bf Source / state} & \jpsi & \psitwos & \jpsi & \psitwos   & \jpsi & \psitwos   & \jpsi & \psitwos   & \jpsi & \psitwos \\  
\midrule 
{\boldmath \bf Mass: PD } &\phz 0.08 &\phz 0.20 &\phz 0.04 &\phz 0.02 &\phz 0.04 &\phz 0.03 &\phz 0.05 &\phz 0.02 &\phz 0.09 &\phz 0.07\\ 
{\boldmath \bf Mass: \psitwos offset } &\phz 0.03 &\phz 0.09 &\phz 0.01 &\phz 0.04 &\phz 0.01 &\phz 0.03 &\phz 0.01 &\phz 0.01 &\phz 0.01 &\phz 0.03\\ 
{\boldmath \bf \ptsq: \chicJ feed-down } &\phz 0.05 &\phz 0.05 &\phz 0.03 &  &\phz 0.07 &  &\phz 0.08 &\phz 0.01 &\phz 0.05 &\phz 0.03\\ 
{\boldmath \bf \ptsq: \psitwos feed-down } &\phz 0.29 &\phz 0.08 &\phz 0.06 &  &  &  &\phz 0.01 &\phz 0.02 &\phz 0.01 &\phz 0.03\\ 
{\boldmath \bf \ptsq: PD shape } &\phz 4.0\phz &\phz 0.9\phz &\phz 0.7\phz &\phz 0.30 &\phz 0.17 &\phz 2.3\phz &\phz 0.03 &\phz 1.0\phz &\phz 0.26 &\phz 0.5\phz\\ 
\midrule 
\boldmath \bf Total systematic &\phz 4.0\phz &\phz 0.9\phz &\phz 0.7\phz &\phz 0.30 &\phz 0.19 &\phz 2.3\phz &\phz 0.10 &\phz 1.0\phz &\phz 0.28 &\phz 0.5\phz\\ 
\midrule 
\boldmath \bf Statistical &\phz 2.6\phz & 14.4\phz &\phz 1.5\phz &\phz 8.0\phz &\phz 1.2\phz &\phz 5.6\phz &\phz 1.0\phz &\phz 5.2\phz &\phz 0.9\phz &\phz 5.0\phz\\ 
\midrule 
\toprule 
{\bf \boldmath $y$  bin} & \multicolumn{2}{c}{\bf 3.25--3.5} & \multicolumn{2}{c}{\bf 3.5--3.75} & \multicolumn{2}{c}{\bf 3.75--4.0} & \multicolumn{2}{c}{\bf 4.0--4.25} & \multicolumn{2}{c}{\bf 4.25--4.5} \\ 
{\bf Source / state} & \jpsi & \psitwos & \jpsi & \psitwos   & \jpsi & \psitwos   & \jpsi & \psitwos   & \jpsi & \psitwos \\  
\midrule 
{\boldmath \bf Mass: PD } &\phz 0.07 &\phz 0.02 &\phz 0.06 &\phz 0.01 &\phz 0.06 &\phz 0.07 &\phz 0.03 &  &\phz 0.04 &\phz 0.25\\ 
{\boldmath \bf Mass: \psitwos offset } &\phz 0.01 &\phz 0.01 &  &\phz 0.03 &  &\phz 0.02 &\phz 0.03 &\phz 0.05 &\phz 0.01 &\phz 0.03\\ 
{\boldmath \bf \ptsq: \chicJ feed-down } &\phz 0.04 &\phz 0.01 &\phz 0.07 &\phz 0.02 &\phz 0.04 &  &\phz 0.03 &\phz 0.01 &\phz 0.01 &\phz 0.03\\ 
{\boldmath \bf \ptsq: \psitwos feed-down } &\phz 0.08 &  &\phz 0.01 &\phz 0.01 &\phz 0.04 &  &\phz 0.10 &\phz 0.03 &\phz 0.04 &\phz 0.05\\ 
{\boldmath \bf \ptsq: PD shape } &\phz 0.17 &\phz 0.26 &\phz 0.13 &\phz 0.6\phz &\phz 0.08 &\phz 1.8\phz &\phz 0.4\phz &\phz 1.7\phz &\phz 2.2\phz &\phz 4.1\phz\\ 
\midrule 
\boldmath \bf Total systematic &\phz 0.20 &\phz 0.26 &\phz 0.16 &\phz 0.6\phz &\phz 0.12 &\phz 1.8\phz &\phz 0.4\phz &\phz 1.7\phz &\phz 2.2\phz &\phz 4.2\phz\\ 
\midrule 
\boldmath \bf Statistical &\phz 0.9\phz &\phz 4.6\phz &\phz 1.0\phz &\phz 5.4\phz &\phz 1.2\phz &\phz 6.3\phz &\phz 1.7\phz &\phz 9.9\phz &\phz 2.9\phz & 20.4\phz\\ 
\midrule 
\bottomrule 
\end{tabular} 

%% file: tables/total_Jpsi2mumu.tex
\sigma_{\jpsi\to\mumu}(2.0<y_{\jpsi}<4.5,2.0<\eta_{\mupm} < 4.5) &= 400 \pm  2 \pm 5 \pm 12 \pb\,,

%% file: tables/total_psi2S2mumu.tex
\sigma_{\psitwos\to\mumu}(2.0<y_{\psitwos}<4.5,2.0<\eta_{\mupm} < 4.5) &= 9.40 \pm  0.15 \pm 0.13 \pm 0.27 \pb\,,

%% file: tables/ratio.tex
\begin{equation*}
	\frac{\sigma_{\psitwos}}{\sigma_{\jpsi}} = 0.1763 \pm 0.0029 \pm 0.0008 \pm 0.0039 \,,
\end{equation*}

%% file: tables/pomeron.tex
\begin{align*} 
  \alpha^{\prime, \jpsi} &= 0.133 \pm 0.024  \pm 0.006  \ (\text{GeV}/c)^{-2},\\ 
  \alpha^{\prime, \psitwos} &= 0.178 \pm 0.124 \pm 0.004\ (\text{GeV}/c)^{-2},\\ 
  b_0^{\jpsi} &= 4.80 \pm 0.24 \pm 0.06 \ (\text{GeV}/c)^{-2},\\ 
  b_0^{\psitwos} &= 4.02 \pm 1.23 \pm 0.03 \ (\text{GeV}/c)^{-2}, 
\end{align*}%

%% file: acknowledgements.tex
\section*{Acknowledgements}
%
%
\noindent We express our gratitude to our colleagues in the CERN
accelerator departments for the excellent performance of the LHC. We
thank the technical and administrative staff at the LHCb
institutes.
We acknowledge support from CERN and from the national agencies:
CAPES, CNPq, FAPERJ and FINEP (Brazil); 
MOST and NSFC (China); 
CNRS/IN2P3 (France); 
BMBF, DFG and MPG (Germany); 
INFN (Italy); 
NWO (Netherlands); 
MNiSW and NCN (Poland); 
MCID/IFA (Romania); 
MICIU and AEI (Spain);
SNSF and SER (Switzerland); 
NASU (Ukraine); 
STFC (United Kingdom); 
DOE NP and NSF (USA).
We acknowledge the computing resources that are provided by CERN, IN2P3
(France), KIT and DESY (Germany), INFN (Italy), SURF (Netherlands),
PIC (Spain), GridPP (United Kingdom), 
CSCS (Switzerland), IFIN-HH (Romania), CBPF (Brazil),
and Polish WLCG (Poland).
We are indebted to the communities behind the multiple open-source
software packages on which we depend.
Individual groups or members have received support from
ARC and ARDC (Australia);
Key Research Program of Frontier Sciences of CAS, CAS PIFI, CAS CCEPP, 
Fundamental Research Funds for the Central Universities, 
and Sci. \& Tech. Program of Guangzhou (China);
Minciencias (Colombia);
EPLANET, Marie Sk\l{}odowska-Curie Actions, ERC and NextGenerationEU (European Union);
A*MIDEX, ANR, IPhU and Labex P2IO, and R\'{e}gion Auvergne-Rh\^{o}ne-Alpes (France);
AvH Foundation (Germany);
ICSC (Italy); 
Severo Ochoa and Mar\'ia de Maeztu Units of Excellence, GVA, XuntaGal, GENCAT, InTalent-Inditex and Prog. ~Atracci\'on Talento CM (Spain);
SRC (Sweden);
the Leverhulme Trust, the Royal Society
 and UKRI (United Kingdom).

%% file: appendix.tex

\appendix
\section*{Appendices}
\section{Photoproduction cross-section}\label{App:photo}

\begin{figure}[p]
	\centering
	\includegraphics[width=0.8\textwidth]{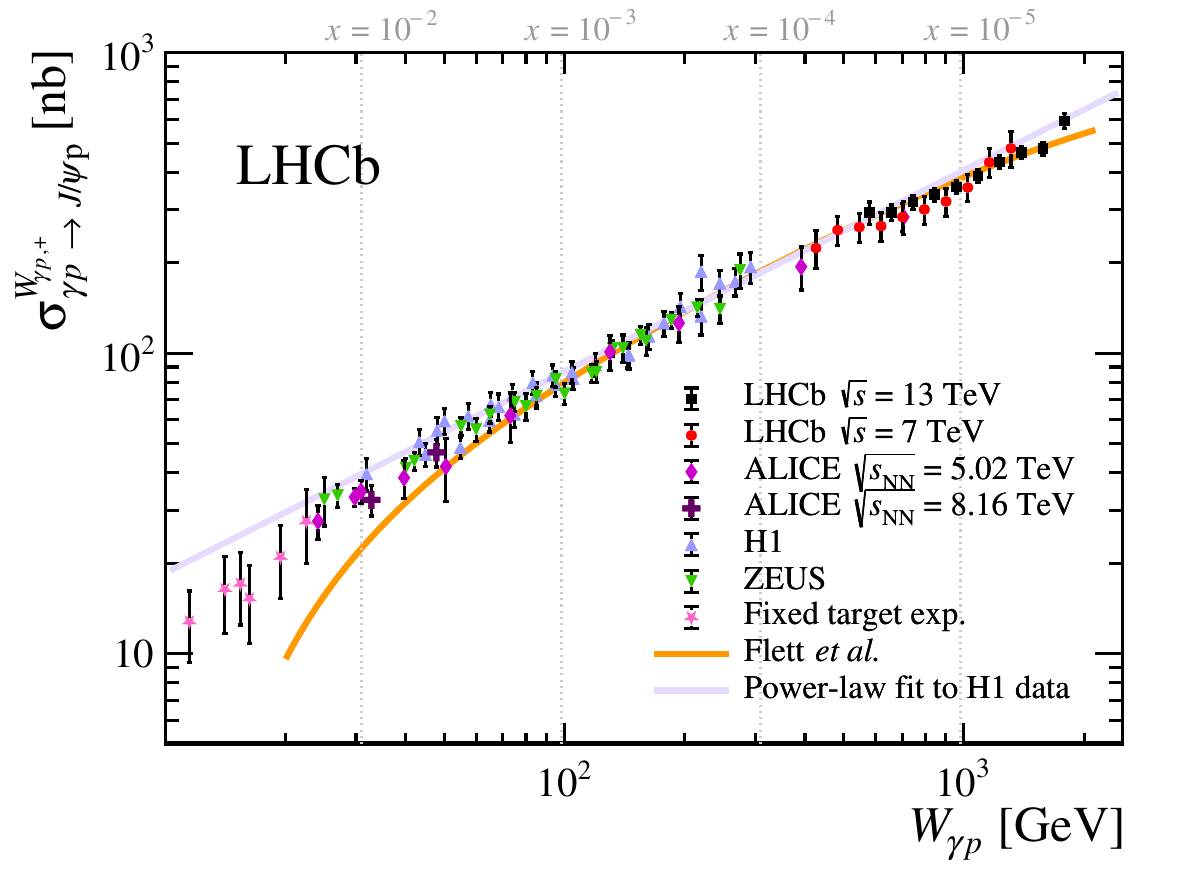}\\
	\caption{Results for the $J\mskip -3mu/\mskip -2mu\psi$ photoproduction cross-section as a function of
          the photon-proton energy $W_{\gamma p}$ from different experiments listed in Table~\ref{tab:photop-refs}.
          The LHCb results at $\sqrt{s} = 13 \tev$ are estimated with the NLO~\cite{Flett:2020duk} calculation.
          Also shown are the NLO theoretical descriptions given by Flett {\it et al.}~\cite{Flett:2020duk},
          as well as a power-law description of the H1 data.
          The top axis shows the values of $x$ reached for a given photon-proton energy.} 
	\label{fig:photoprod-FMRT}
\end{figure}

The differential cross-sections are used to determine the photoproduction
cross-section for \jpsi and \psitwos mesons.
The differential cross-section is factorised into two terms depending on whether the proton
travelling from the vertex detector towards the
muon chambers interacts electromagnetically (labelled $W_{\gamma p,+}$) or the opposite-direction proton does (labelled $W_{\gamma p,-}$):
\begin{equation}
	\frac{{\rm d}\sigma}{{\rm d}y}(pp \rightarrow p\psi p) = S^2(W_{\gamma p,+}) \left( k_+ \frac{{\rm d}n}{{\rm d}k_+} \right) \sigma_{\gamma p \rightarrow \psi p}^{W_{\gamma p,+}} + S^2(W_{\gamma p,-}) \left( k_- \frac{{\rm d}n}{{\rm d}k_-} \right)  \sigma_{\gamma p \rightarrow \psi p}^{W_{\gamma p,-}}, \label{eq:photoprod-Y}
\end{equation}
with $W_{\gamma p, \pm} = \sqrt{M_{\psi}c^2 \sqrt{s} e^{\pm|y|}}$.
The $S^2(W_{\gamma p,\pm})$ terms, the so-called survival factors, are taken from Ref.~\cite{Jones:2016icr}.
The photon flux ${\rm d}n/{\rm d}k_{\pm}$ for photons with energy equal to \mbox{$k_{\pm} = (M_{\psi}c^2 /2)e^{\pm|y|}$} is calculated following Refs.~\cite{Budnev:1975poe,Kepka:1255854}.
The photoproduction cross-sections are given by $\sigma_{\gamma p \rightarrow \psi p}^{W_{\gamma p,\pm}}$.
The antiparallel $\gamma p$ cross-section, $\sigma_{\gamma p \rightarrow \psi p}^{W_{\gamma p,-}}$, corresponds to large values of $x$, as \mbox{$x \sim M_{\psi}c^2 /\sqrt{s}\,e^{-y}$}~\cite{Flett:2020duk}.
The contribution of this term to Eq.~\ref{eq:photoprod-Y} is therefore expected to be small and can be constrained from theoretical predictions.
The antiparallel solution is taken from the \jpsi and \psitwos NLO cross-section predictions from Refs.~\cite{Flett:2020duk,Jones:2013eda}
and subtracted.
Figure~\ref{fig:photoprod-FMRT} shows the measured photoproduction cross-section for \jpsi mesons
and compares it with previous measurements listed in Table~\ref{tab:photop-refs}.
The numerical values are listed in Table~\ref{tab:photop-Jpsi}.
\begin{table}[p]
	\centering
	\caption{Previous results used in Fig.~\ref{fig:photoprod-FMRT}.}\label{tab:photop-refs}
  \input{tables/photoproduction}
\end{table}

The LHCb \jpsi data at $\sqrt{s} = 13 \tev$ are in agreement with the NLO description~\cite{Flett:2020duk}.
They also follow a power-law fit to H1 data~\cite{H1:2013okq}.


\FloatBarrier
\section{Numerical results}\label{App:numbers}

Tables~\ref{tab:crossSection-theory-Jpsi2mumu} and~\ref{tab:crossSection-theory-Psi2S2mumu} present the differential cross-sections in rapidity bins, shown in Fig.~\ref{fig:crossSectionMM}, and the breakdown of uncertainties for \jpmm and \psimm, respectively. Their ratio is given in Table~\ref{tab:ratio}.
%
\begin{table}[b]
	\caption{Differential CEP \jpmm yields and cross-sections corrected for efficiency ($\epsilon_{{\rm tot}}$), acceptance ($\epsilon_{\rm Geom.Acc.}$) and branching fraction. The systematic uncertainties are split between those uncorrelated across $y$ ranges, those that are 100\% correlated and the luminosity uncertainty.} \label{tab:crossSection-theory-Jpsi2mumu}
	\resizebox{\textwidth}{!}{
          \input{tables/tableLaTeX_crossSection_Jpsi2mumu_PAPER}
	}
\end{table}
\begin{table}[h]
	\caption{Differential CEP \psimm yields and cross-sections corrected for efficiency ($\epsilon_{{\rm tot}}$), acceptance ($\epsilon_{\rm Geom.Acc.}$) and branching fraction. The systematic uncertainties are split between those uncorrelated across $y$ ranges, those that are 100\% correlated and the luminosity uncertainty.} \label{tab:crossSection-theory-Psi2S2mumu}
	\resizebox{\textwidth}{!}{
          \input{tables/tableLaTeX_crossSection_psi2S2mumu_PAPER}
	}
\end{table}
%
\begin{table}[tb]
  \caption{Ratio of the CEP $J\mskip -3mu/\mskip -2mu\psi\!\to\mu^+\mu^-$ and $\psi(2S)\!\to\mu^+\mu^-$ cross-sections per rapidity bin.}\label{tab:ratio}
  \input{tables/tableLaTeX_ratio}
\end{table}
%
Table~\ref{tab:bSlopes} lists the exponential slopes for \jpsi and \psitwos in each rapidity bin, corresponding to Fig.~\ref{fig:shrinkageFit}. 
%
\begin{table}[t]\centering
  \caption{Values of the $b$ slopes measured in the fit.}\label{tab:bSlopes}
  \input{tables/tableLaTeX_bSlopes.tex}
\end{table}
%
Table~\ref{tab:photop-Jpsi} lists the values entering the computation of the parallel cross-sections
$\sigma_{\gamma p \rightarrow \jpsi p}^{W_{\gamma p,+}}$ displayed in Fig.~\ref{fig:photoprod-FMRT}.
%
\begin{table}[tb]\centering
  \caption{Values of the differential cross-section, survival factor, photon flux and antiparallel photoproduction cross-section used for the calculation of the parallel cross-section. The antiparallel $\gamma p$ value  is taken from the FMRT NLO description~\cite{Flett:2020duk}.}\label{tab:photop-Jpsi}
  \input{tables/tableLaTeX_photoproduction_crossSection_Jpsi2mumu_FMRT}
\end{table}

\FloatBarrier
\cleardoublepage
\section{Fits in all rapidity bins}\label{App:fits}
The distributions of dimuon mass and $p_{\mathrm{T}}^2$ in each rapidity interval are shown in Fig.~\ref{fig:systUnc_ptsq_to_5_0-both_PDH1}. The results of the two-dimensional fits described in the main text are overlaid. 

\SystematicsPlots{figs/Fig11}{ptsq_to_5_0-both}{PDH1}  
\FloatBarrier

%% file: tables/photoproduction.tex
\begin{tabular}{c|l|cc|c} 
  \toprule
  Marker & Experiment & collision & Energy & Refs. \\
\midrule 
  {\color{red} $\bullet$} & LHCb & $pp$ & $\sqrt{s}=7\tev$ & \cite{LHCb-PAPER-2013-059} \\
  {\color{violet} $\blacklozenge$} & ALICE & $p$Pb & $\sqsnn = 5.02 \tev$ & \cite{ALICE:2014eof,ALICE:2018oyo} \\
  {\color{violet} $\boldsymbol +$} & ALICE & $p$Pb & $\sqsnn = 8.16 \tev$~ & \cite{ALICE:2023mfc} \\
  {\color{blue} $\blacktriangle$} & H1 & $ep$ & $40 < W_{\gamma p} < 305 \gev$ & \cite{H1:2005dtp} \\
  {\color{blue} $\blacktriangle$} & H1 & $ep$ & $25 < W_{\gamma p} < 110 \gev$ & \cite{H1:2013okq} \\
  {\color{green} $\blacktriangledown$} & ZEUS & $ep$ & $20 < W_{\gamma p} < 290 \gev$ & \cite{ZEUS:2002wfj} \\
  {\color{pink} \rotatebox[origin=c]{180}{$\boldsymbol\star$}} & E87 & $\gamma$Be & $0<E_\gamma<250\gev$ & \cite{Knapp:1975nf} \\
  {\color{pink} \rotatebox[origin=c]{180}{$\boldsymbol\star$}} & E401 & $\gamma$H and $\gamma{}^2$H & $60<E_\gamma<300\gev$ &  \cite{Binkley:1981kv} \\
  {\color{pink} \rotatebox[origin=c]{180}{$\boldsymbol\star$}} & E516 & $\gamma$H & $60<E_\gamma<160\gev$ &  \cite{Denby:1983az} \\
\bottomrule 
\end{tabular}

%% file: tables/tableLaTeX_crossSection_Jpsi2mumu_PAPER.tex
\begin{tabular}{lccccc} 
\toprule 
{\boldmath $y_{\jpsi}$ {\bf bin}} & {\bf 2.0--2.25} & {\bf 2.25--2.5} & {\bf 2.5--2.75} & {\bf 2.75--3.0} & {\bf 3.0--3.25} \\ 
\midrule 
{\boldmath $N_{{\rm sig}}$}  & $4998 \pm 113\phz$  & $18095 \pm 265\phz\phz$  & $31591 \pm 361\phz\phz$  & $41640 \pm 402\phz\phz$  & $47690 \pm 432\phz\phz$  \\ 
{\boldmath $\epsilon_{{\rm tot}}$} & $0.313 \pm 0.028$ & $0.403 \pm 0.030$ & $0.443 \pm 0.030$ & $0.456 \pm 0.029$ & $0.467 \pm 0.027$\\
{\boldmath $\epsilon_{\rm Geom.Acc.}$} & $0.095 \pm 0.002$ & $0.287 \pm 0.003$ & $0.466 \pm 0.003$ & $0.623 \pm 0.003$ & $0.732 \pm 0.003$\\
{\boldmath \bf ${{\rm d}}\sigma/{{\rm d}}y$ [nb]} & $7.41$ & $6.90$ & $6.75$ & $6.46$ & $6.15$\\
{\boldmath \bf Stat. unc. [nb]} & $0.17$ & $0.10$ & $0.08$ & $0.06$ & $0.06$\\
{\boldmath \bf Uncorr. syst. unc. [nb]} & $0.30$ & $0.12$ & $0.11$ & $0.10$ & $0.09$\\
{\boldmath \bf Corr. syst. unc. [nb]} & $0.23$ & $0.21$ & $0.21$ & $0.20$ & $0.19$\\
{\boldmath \bf Lumi. unc. [nb]} & $0.21$ & $0.20$ & $0.20$ & $0.19$ & $0.18$\\
\bottomrule 
\toprule 
{\boldmath $y_{\jpsi}$ {\bf bin}} & {\bf 3.25--3.5} & {\bf 3.5--3.75} & {\bf 3.75--4.0} & {\bf 4.0--4.25} & {\bf 4.25--4.5} \\ 
\midrule 
{\boldmath $N_{{\rm sig}}$}  & $47303 \pm 436\phz\phz$  & $39878 \pm 394\phz\phz$  & $26727 \pm 329\phz\phz$  & $14428 \pm 236\phz\phz$  & $4349 \pm 108\phz$  \\ 
{\boldmath $\epsilon_{{\rm tot}}$} & $0.479 \pm 0.027$ & $0.484 \pm 0.028$ & $0.462 \pm 0.029$ & $0.435 \pm 0.029$ & $0.399 \pm 0.029$\\
{\boldmath $\epsilon_{\rm Geom.Acc.}$} & $0.733 \pm 0.003$ & $0.625 \pm 0.003$ & $0.467 \pm 0.003$ & $0.300 \pm 0.003$ & $0.095 \pm 0.002$\\
{\boldmath \bf ${{\rm d}}\sigma/{{\rm d}}y$ [nb]} & $5.93$ & $5.82$ & $5.47$ & $4.89$ & $5.05$\\
{\boldmath \bf Stat. unc. [nb]} & $0.05$ & $0.06$ & $0.07$ & $0.08$ & $0.13$\\
{\boldmath \bf Uncorr. syst. unc. [nb]} & $0.09$ & $0.10$ & $0.10$ & $0.11$ & $0.12$\\
{\boldmath \bf Corr. syst. unc. [nb]} & $0.18$ & $0.18$ & $0.17$ & $0.15$ & $0.16$\\
{\boldmath \bf Lumi. unc. [nb]} & $0.17$ & $0.17$ & $0.16$ & $0.14$ & $0.15$\\
\bottomrule 
\end{tabular} 

%% file: tables/tableLaTeX_crossSection_psi2S2mumu_PAPER.tex
\begin{tabular}{lccccc} 
\toprule 
{\boldmath $y_{\psitwos}$ {\bf bin}} & {\bf 2.0--2.25} & {\bf 2.25--2.5} & {\bf 2.5--2.75} & {\bf 2.75--3.0} & {\bf 3.0--3.25} \\ 
\midrule 
{\boldmath $N_{{\rm sig}}$}  & $127 \pm 16\phz$  & $491 \pm 30\phz$  & $845 \pm 39\phz$  & $1088 \pm 44\phz\phz$  & $1163 \pm 47\phz\phz$  \\ 
{\boldmath $\epsilon_{{\rm tot}}$} & $0.400 \pm 0.032$ & $0.494 \pm 0.032$ & $0.527 \pm 0.032$ & $0.518 \pm 0.032$ & $0.502 \pm 0.031$\\
{\boldmath $\epsilon_{\rm Geom.Acc.}$} & $0.091 \pm 0.002$ & $0.284 \pm 0.003$ & $0.465 \pm 0.003$ & $0.626 \pm 0.003$ & $0.735 \pm 0.003$\\
{\boldmath \bf ${{\rm d}}\sigma/{{\rm d}}y$ [nb]} & $1.16$ & $1.16$ & $1.14$ & $1.11$ & $1.05$\\
{\boldmath \bf Stat. unc. [nb]} & $0.15$ & $0.07$ & $0.05$ & $0.05$ & $0.04$\\
{\boldmath \bf Uncorr. syst. unc. [nb]} & $0.03$ & $0.02$ & $0.02$ & $0.02$ & $0.02$\\
{\boldmath \bf Corr. syst. unc. [nb]} & $0.04$ & $0.04$ & $0.04$ & $0.03$ & $0.03$\\
{\boldmath \bf Lumi. unc. [nb]} & $0.03$ & $0.03$ & $0.03$ & $0.03$ & $0.03$\\
\bottomrule 
\toprule 
{\boldmath $y_{\psitwos}$ {\bf bin}} & {\bf 3.25--3.5} & {\bf 3.5--3.75} & {\bf 3.75--4.0} & {\bf 4.0--4.25} & {\bf 4.25--4.5} \\ 
\midrule 
{\boldmath $N_{{\rm sig}}$}  & $1319 \pm 48\phz\phz$  & $980 \pm 42\phz$  & $648 \pm 34\phz$  & $323 \pm 25\phz$  & $81 \pm 13$  \\ 
{\boldmath $\epsilon_{{\rm tot}}$} & $0.505 \pm 0.030$ & $0.492 \pm 0.030$ & $0.457 \pm 0.030$ & $0.427 \pm 0.030$ & $0.394 \pm 0.035$\\
{\boldmath $\epsilon_{\rm Geom.Acc.}$} & $0.740 \pm 0.003$ & $0.622 \pm 0.003$ & $0.469 \pm 0.003$ & $0.290 \pm 0.003$ & $0.099 \pm 0.002$\\
{\boldmath \bf ${{\rm d}}\sigma/{{\rm d}}y$ [nb]} & $1.17$ & $1.06$ & $1.00$ & $0.86$ & $0.69$\\
{\boldmath \bf Stat. unc. [nb]} & $0.04$ & $0.05$ & $0.05$ & $0.07$ & $0.11$\\
{\boldmath \bf Uncorr. syst. unc. [nb]} & $0.02$ & $0.02$ & $0.03$ & $0.02$ & $0.02$\\
{\boldmath \bf Corr. syst. unc. [nb]} & $0.04$ & $0.03$ & $0.03$ & $0.03$ & $0.02$\\
{\boldmath \bf Lumi. unc. [nb]} & $0.03$ & $0.03$ & $0.03$ & $0.02$ & $0.02$\\
\bottomrule 
\end{tabular} 

%% file: tables/tableLaTeX_ratio.tex
\begin{center} 
\begin{tabular}{lccccc} 
\toprule 
{\boldmath $y$ {\bf bin}} & {\bf 2.0--2.25} & {\bf 2.25--2.5} & {\bf 2.5--2.75} & {\bf 2.75--3.0} & {\bf 3.0--3.25} \\ 
\midrule 
{\boldmath $\frac{{\rm d}\sigma_{\psitwos}/{\rm d}y}{{\rm d}\sigma_{\jpsi}/{\rm d}y}$} & $0.156$ & $0.168$ & $0.169$ & $0.172$ & $0.170$\\
{\boldmath \bf Stat. unc.} & $0.020$ & $0.010$ & $0.008$ & $0.007$ & $0.007$\\
{\boldmath \bf Syst. unc.} & $0.006$ & $0.001$ & $0.001$ & $0.001$ & $0.002$\\
{\boldmath \bf BF unc.} & $0.003$ & $0.004$ & $0.004$ & $0.004$ & $0.004$\\
\bottomrule 
\toprule 
{\boldmath $y$ {\bf bin}} & {\bf 3.25--3.5} & {\bf 3.5--3.75} & {\bf 3.75--4.0} & {\bf 4.0--4.25} & {\bf 4.25--4.5} \\ 
\midrule 
{\boldmath $\frac{{\rm d}\sigma_{\psitwos}/{\rm d}y}{{\rm d}\sigma_{\jpsi}/{\rm d}y}$} & $0.197$ & $0.182$ & $0.183$ & $0.177$ & $0.137$\\
{\boldmath \bf Stat. unc.} & $0.007$ & $0.008$ & $0.010$ & $0.014$ & $0.022$\\
{\boldmath \bf Syst. unc.} & $0.002$ & $0.003$ & $0.004$ & $0.005$ & $0.003$\\
{\boldmath \bf BF unc.} & $0.004$ & $0.004$ & $0.004$ & $0.004$ & $0.003$\\
\bottomrule 
\end{tabular} 
\end{center} 

%% file: tables/tableLaTeX_bSlopes.tex
\begin{tabular}{c|cc|cc} 
\toprule 
{\boldmath $y$} & {\boldmath $\log\left(\frac{W^{\jpsi}_{\gamma p}}{W_0}\right)$} & {\boldmath $b^\jpsi$} & {\boldmath $\log\left(\frac{W^{\psitwos}_{\gamma p}}{W_0}\right)$} & {\boldmath $b^\psitwos$} \\ 
\midrule 
{\bf 2.0--2.25 } & $1.86$ & $6.07 \pm 0.16 \pm 0.25$ & $1.95$ & $6.1 \pm 0.9 \pm 0.1$ \\ 
{\bf 2.25--2.5 } & $1.99$ & $5.83 \pm 0.09 \pm 0.04$ & $2.08$ & $5.4 \pm 0.4 \pm 0.0$ \\ 
{\bf 2.5--2.75 } & $2.11$ & $5.90 \pm 0.07 \pm 0.01$ & $2.20$ & $5.7 \pm 0.3 \pm 0.1$ \\ 
{\bf 2.75--3.0 } & $2.24$ & $5.99 \pm 0.06 \pm 0.01$ & $2.33$ & $5.7 \pm 0.3 \pm 0.1$ \\ 
{\bf 3.0--3.25 } & $2.36$ & $6.02 \pm 0.05 \pm 0.02$ & $2.45$ & $5.9 \pm 0.3 \pm 0.0$ \\ 
{\bf 3.25--3.5 } & $2.49$ & $6.25 \pm 0.06 \pm 0.01$ & $2.58$ & $5.5 \pm 0.2 \pm 0.0$ \\ 
{\bf 3.5--3.75 } & $2.61$ & $6.13 \pm 0.06 \pm 0.01$ & $2.70$ & $5.7 \pm 0.3 \pm 0.0$ \\ 
{\bf 3.75--4.0 } & $2.74$ & $6.20 \pm 0.08 \pm 0.01$ & $2.83$ & $6.8 \pm 0.4 \pm 0.1$ \\ 
{\bf 4.0--4.25 } & $2.86$ & $6.33 \pm 0.11 \pm 0.02$ & $2.95$ & $6.3 \pm 0.6 \pm 0.1$ \\ 
{\bf 4.25--4.5 } & $2.99$ & $6.51 \pm 0.19 \pm 0.14$ & $3.08$ & $7.9 \pm 1.6 \pm 0.3$ \\ 
\bottomrule 
\end{tabular} 

%% file: tables/tableLaTeX_photoproduction_crossSection_Jpsi2mumu_FMRT.tex
\begin{tabular}{lccccc} 
\toprule 
{\boldmath $y_{\jpsi}$} {\bf bin} & {\bf 2.0--2.25} & {\bf 2.25--2.5} & {\bf 2.5--2.75} & {\bf 2.75--3.0} & {\bf 3.0--3.25} \\ 
\midrule 
{\boldmath \bf ${\rm d}\sigma / {\rm d}y$ [nb]} & $7.41 \pm 0.43$ & $6.90 \pm 0.27$ & $6.75 \pm 0.25$ & $6.46 \pm 0.24$ & $6.15 \pm 0.22$\\[0.2cm]
{\boldmath $S^2(W_{\gamma p,+})$} & $0.786$ & $0.774$ & $0.762$ & $0.748$ & $0.732$\\[0.2cm]
{\boldmath $k_{+}\frac{{\rm d}n}{{\rm d}k_{+}} (\times 10^{-3})$} & $22.7$ & $21.6$ & $20.4$ & $19.2$ & $18.0$\\[0.2cm]
{\boldmath $S^2(W_{\gamma p,-})$} & $0.885$ & $0.888$ & $0.891$ & $0.893$ & $0.896$\\[0.2cm]
{\boldmath $k_{-}\frac{{\rm d}n}{{\rm d}k_-} (\times 10^{-3})$} & $42.5$ & $43.7$ & $44.9$ & $46.0$ & $47.2$\\[0.2cm]
{\boldmath \bf $\sigma_{\gamma p \rightarrow \jpsi p}^{W_{\gamma p,-}}$ [nb]} & $57.7 \pm 2.7$ & $51.1 \pm 2.3$ & $45.0 \pm 2.0$ & $39.2 \pm 1.7$ & $33.9 \pm 1.4$\\[0.2cm]
{\boldmath \bf $\sigma_{\gamma p \rightarrow \jpsi p}^{W_{\gamma p,+}}$ [nb]} & $294 \pm 25$ & $294 \pm 17$ & $319 \pm 17$ & $337 \pm 17$ & $358 \pm 17$\\[0.2cm]
\toprule 
\bottomrule 
{\boldmath $y_{\jpsi}$} {\bf bin} & {\bf 3.25--3.5} & {\bf 3.5--3.75} & {\bf 3.75--4.0} & {\bf 4.0--4.25} & {\bf 4.25--4.5} \\ 
\midrule 
{\boldmath \bf ${\rm d}\sigma / {\rm d}y$ [nb]} & $5.93 \pm 0.21$ & $5.82 \pm 0.21$ & $5.47 \pm 0.21$ & $4.89 \pm 0.21$ & $5.05 \pm 0.26$\\[0.2cm]
{\boldmath $S^2(W_{\gamma p,+})$} & $0.715$ & $0.695$ & $0.672$ & $0.647$ & $0.618$\\[0.2cm]
{\boldmath $k_{+}\frac{{\rm d}n}{{\rm d}k_{+}} (\times 10^{-3})$} & $16.8$ & $21.6$ & $14.5$ & $13.3$ & $12.1$\\[0.2cm]
{\boldmath $S^2(W_{\gamma p,-})$} & $0.899$ & $0.901$ & $0.903$ & $0.905$ & $0.907$\\[0.2cm]
{\boldmath $k_{-}\frac{{\rm d}n}{{\rm d}k_-} (\times 10^{-3})$} & $48.3$ & $49.5$ & $50.7$ & $51.8$ & $53.0$\\[0.2cm]
{\boldmath \bf $\sigma_{\gamma p \rightarrow \jpsi p}^{W_{\gamma p,-}}$ [nb]} & $29.0 \pm 1.2$ & $24.4 \pm 0.9$ & $20.1 \pm 0.7$ & $16.2 \pm 0.6$ & $12.6 \pm 0.4$\\[0.2cm]
{\boldmath \bf $\sigma_{\gamma p \rightarrow \jpsi p}^{W_{\gamma p,+}}$ [nb]} & $389 \pm 18$ & $433 \pm 20$ & $467 \pm 22$ & $480 \pm 25$ & $594 \pm 34$\\[0.2cm]
\toprule 
\bottomrule 
\end{tabular} 

%% file: Authorship_LHCb-PAPER-2024-012.tex
\centerline
{\large\bf LHCb collaboration}
\begin
{flushleft}
\small
R.~Aaij$^{36}$\lhcborcid{0000-0003-0533-1952},
A.S.W.~Abdelmotteleb$^{55}$\lhcborcid{0000-0001-7905-0542},
C.~Abellan~Beteta$^{49}$,
F.~Abudin{\'e}n$^{55}$\lhcborcid{0000-0002-6737-3528},
T.~Ackernley$^{59}$\lhcborcid{0000-0002-5951-3498},
A. A. ~Adefisoye$^{67}$\lhcborcid{0000-0003-2448-1550},
B.~Adeva$^{45}$\lhcborcid{0000-0001-9756-3712},
M.~Adinolfi$^{53}$\lhcborcid{0000-0002-1326-1264},
P.~Adlarson$^{79}$\lhcborcid{0000-0001-6280-3851},
C.~Agapopoulou$^{13}$\lhcborcid{0000-0002-2368-0147}
% ...

{\footnotesize \it

% ...
$^{13}$Universit{\'e} Paris-Saclay, CNRS/IN2P3, IJCLab, Orsay, France\\
% ...
$^{36}$Nikhef National Institute for Subatomic Physics, Amsterdam, Netherlands\\
% ...
$^{45}$Instituto Galego de F{\'\i}sica de Altas Enerx{\'\i}as (IGFAE), Universidade de Santiago de Compostela, Santiago de Compostela, Spain\\
% ...
$^{49}$Physik-Institut, Universit{\"a}t Z{\"u}rich, Z{\"u}rich, Switzerland\\
% ...
$^{53}$H.H. Wills Physics Laboratory, University of Bristol, Bristol, United Kingdom\\
% ...
$^{55}$Department of Physics, University of Warwick, Coventry, United Kingdom\\
% ...
$^{58}$School of Physics and Astronomy, University of Glasgow, Glasgow, United Kingdom\\
$^{59}$Oliver Lodge Laboratory, University of Liverpool, Liverpool, United Kingdom\\
% ...
$^{67}$Syracuse University, Syracuse, NY, United States\\
% ...
$^{79}$Department of Physics and Astronomy, Uppsala University, Uppsala, Sweden, associated to $^{58}$\\
% ...
}
\end{flushleft}